\documentclass[11pt,a4paper]{article}
\usepackage[utf8]{inputenc}
\usepackage[english]{babel}
\usepackage[a4paper, total={6in, 9in}]{geometry}
\usepackage{amsfonts}
\usepackage{amsmath}
\usepackage{amssymb}
\usepackage{graphicx}
\usepackage{slashed}
\usepackage{array}
\usepackage{mathrsfs}
\usepackage{float}
\restylefloat{table}
\usepackage{verbatim}
\usepackage{xcolor}
\usepackage{upgreek}
\usepackage{comment}
\usepackage{multirow}
\usepackage{enumitem}
\usepackage{mathtools}
\usepackage[vcentermath]{youngtab}
\usepackage{jheppub}

\usepackage{multirow}

\usepackage[T1]{fontenc} 
\include{amsmath}
\usepackage{subcaption}

\usepackage[compat=1.1.0]{tikz-feynhand}
\usepackage{tikz-feynman}
\tikzfeynmanset{compat=1.1.0}
\usepackage{feynmp}
\usepackage{tikzsymbols}
\usepackage{array}
\usepackage{pifont} 

\newcommand{\dg}{^{\dagger}}
\newcommand{\hc}{\text{h.c.}}

\newcommand{\sun}{\mathrm{SU}(N)_{\mathrm{DC}}}

\newcommand{\sut}{\mathrm{SU}(3)_{\mathrm{DC}}}

\newcommand{\ldc}{\Lambda_{\mathrm{DC}}}
\newcommand{\ndc}{N_{\mathrm{DC}}}
\newcommand{\mtcb}{m_{\mathrm{DCb}}}
\newcommand{\Tr}{\mathrm{Tr}}

\newcommand{\dPi}{\mathrm{d}\Pi}
\newcommand{\sigmaone}{\langle\sigma_1 v\rangle}

\newcommand{\luv}{\Lambda_{\mathrm{UV}}}
\newcommand{\lcut}{\Lambda_{\mathrm{cut}}}
\newcommand{\udb}{\mathrm{U}(1)_{\mathrm{DB}}}
\newcommand{\tcp}{DC$\pi~$}
\newcommand{\hub}{\mathrm{H}}

\title{Asymmetric accidental composite dark matter}
\date{}
\author[a,b]{Salvatore Bottaro,}
\author[a,b]{Marco Costa,}
\author[a,c]{Oleg Popov}
\affiliation[a]{Scuola Normale Superiore, Piazza dei Cavalieri 7, 56126 Pisa, Italy}
\affiliation[b]{INFN Sezione di Pisa, Largo B. Pontecorvo 3, I-56127 Pisa, Italy}
\affiliation[c]{Department of Physics, Korea Advanced Institute of Science and Technology, 291 Daehak-ro, Yuseong-gu, Daejeon 34141, Republic of Korea}
\emailAdd{salvatore.bottaro@sns.it}
\emailAdd{marco.costa@sns.it}
\emailAdd{oleg.popov@sns.it}

\abstract{The goal of this work is to find the simplest UV completion of Accidental Composite Dark Matter Models (ACDM) that can dynamically generate an asymmetry for the DM candidate, the lightest \emph{dark baryon} (DCb), and simultaneously annihilate the symmetric component. In this framework the DCb is a bound state of a confining $\sun$ gauge group, and can interact weakly with the visible sector. The constituents of the DCb can possess non-trivial charges under the Standard Model gauge group.
The generation of asymmetry for such candidate is a two-flavor variation of the \emph{out-of-equilibrium} decay of a heavy scalar, with mass $M_\phi\gtrsim 10^{10}$ GeV. Below the scale of the scalars, the models recover accidental stability, or long-livedness, of the DM candidate. The symmetric component is annihilated by residual confined interactions provided that the mass of the DCb $\mtcb \lesssim 75$ TeV. We implement the mechanism of asymmetry generation, or a variation of it, in all the original ACDM models, managing to generate the correct asymmetry for DCb of masses in this range. For some of the models found, the stability of the DM candidate is not spoiled even considering generic GUT completions or asymmetry generation mechanisms in the visible sector.}

\begin{document}
\maketitle

\section{Introduction}
The nature of Dark Matter (DM) is unknown, and its understanding is one of the major problems of fundamental physics. Despite the huge experimental effort, not even its interactions with the Standard Model (SM) particles are known, except for the fact that are very weak. There is a vast literature of particle physics models that tries to explain the DM origin.
A possible way to characterize a DM model is according to the properties of its DM candidate.
One of the major features of the dark sector (DS) is that it can either be made by a symmetric abundance of dark matter particles and antiparticles, like in typical \emph{weakly interacting massive particles} (WIMP) models. The other popular possibility is that it is \emph{asymmetric} (ADM), being made only by particles and not their charge-conjugates. This is very similar to what happens in the visible universe, that possesses a matter-antimatter asymmetry (BAU) $\eta_b \equiv n_b/s\sim 10^{-10}$.
If the DM is asymmetric, there are interesting differences with the symmetric scenario, regarding for example indirect detection (ID) bounds \cite{10.21468/SciPostPhys.4.6.041,Mahbubani:2019pij}, evolution of astrophysical objects \cite{PhysRevD.87.123507,Kouvaris:2010jy}, and annihilation cross section predictions \cite{Graesser:2011wi}. Therefore this property is not a mere academic curiosity, rather it bears a phenomenologically distinct scenario.\\
Experimentally, the dark matter abundance is roughly of the same order of the baryonic (visible) abundance \cite{Aghanim:2018eyx, Ade:2013zuv}:
\begin{equation}\label{eq:omegarel}
\Omega_{\mathrm{DM}}\simeq 5\Omega _{\mathrm{b}} \;.
\end{equation}
This numerical coincidence, in conjunction with the existence of BAU in the visible sector, has led to many speculations about a possible common origin for the DM and visible sector abundances (see \cite{Petraki:2013wwa, Zurek_2014} for reviews).
In particular, if the DM is asymmetric, the ratio between the two energy densities can be explained by building models that predict a $\mathcal{O}(1)$ relation between the asymmetry in the DS and in the visible sector. The relation \ref{eq:omegarel} is then obtained by taking the mass of the DM to be in the $1\div 10$ GeV.
Typically in this class of models only the numerical density coincidence is explained, while the explanation for having a DM mass near the proton mass (needed to enforce equation \ref{eq:omegarel}) is not given \footnote{See \cite{Lonsdale:2018xwd} for an exception based on \emph{mirror world} framework.}.
The task is usually accomplished by transferring some primordial asymmetry between the visible and dark sector, for example through higher dimensional operators \cite{PhysRevD.79.115016}, or through anomalous electroweak (EW) interactions like \emph{sphalerons} \cite{PhysRevD.44.3062}.
The primordial asymmetry can be generated in several ways. For example it can be generated in the visible sector via the decay of some heavy states (like in \emph{leptogenesis} \cite{Fukugita:1986hr} or \emph{GUT baryogenesis} \cite{PhysRevD.20.2484}), in the dark sector first via decays\cite{Haba:2010bm} or dark first order phase transitions \cite{Shelton:2010ta, DUTTA2011364, Hall:2019rld}, or simultaneously in both sectors from the decay of a heavy particle \cite{PhysRevD.71.023510}.\\
Another distinctive trait of the DM is that it can be made by elementary or \emph{composite} particles \cite{Kribs:2016cew}.
The prototypical composite scenario describes some constituent particles, that we will improperly label \emph{dark quarks}, that are bound together via a new confining gauge interaction, called \emph{dark color}, in baryon-like or pion-like bound states (respectively \emph{dark baryons} and \emph{dark pions}).
The dark quarks can also carry non-trivial charges under the SM gauge group.
The confinement of the dark quarks inside a dark color singlet bound state can be exploited to conceal their SM charges inside a globally SM-neutral (or weakly interacting) bound state \cite{Antipin:2015xia, Appelquist:2015yfa}, if the charges of the constituents are properly chosen. In this way bounds from direct and indirect detection can be evaded, but the presence of potentially SM-charged resonances leads to peculiar collider signatures, that can be tested at current or future experiments.
Another consequence of compositeness is that dark color gauge invariance can lead to a residual symmetry that protects the DM candidate from decaying, without the need to impose by hand further global or discrete symmetries. Combining these two concepts is the idea behind \emph{accidental composite dark matter} (ACDM) models, that were fully classified in \cite{Antipin:2015xia}.\\
Composite asymmetric dark matter models were previously explored in the literature since the early days of \emph{technicolor} (TC) theories \cite{NUSSINOV198555, BARR1990387}.
Unlike ACDM models, the field content of original TC theories is chiral and the mass of the composite technibaryon is naturally tied to the weak scale, given that the goal of such theories is to dynamically generate it.
The chiral field content and the scale coincidence allows the possibility of having the transfer mechanism, the electroweak sphaleron \cite{KUZMIN198536}, to decouple at temperatures at which the DM candidate starts to become non-relativistic. As a consequence, a Boltzmann suppression factor generates a hierarchy between the DM and visible number density asymmetries, and therefore a hierarchy in the masses needed to explain the DM abundance.
Interestingly enough, the mass obtained from reproducing the observed DM relic density naturally falls into the TeV range, as predicted by TC theories \cite{Gudnason:2006yj, Ryttov:2008xe,Frandsen:2009mi, Khlopov:2007ic} (see \cite{Kribs:2009fy} for a non-TC model in which the same happens)\footnote{In a certain sense, this class of TC-based models do not have a $\mathcal{O}(1)$ relation between the number densities, but naturally explain the relation in equation \ref{eq:omegarel} by compensating an exponentially suppressed number density with a hierarchically larger mass of the technibaryon with respect to the proton mass.}. 
Other composite asymmetric dark matter models instead do not link the scale of the DM with the EW scale. Typically they follow the route of predicting a $\mathcal{O}(1)$ relation between the DS and SM number densities, which implies $m_{\mathrm{DM}}\sim \mathcal{O}(1)$ GeV \cite{Ibe:2018juk, Ibe:2018tex, Fukuda:2014xqa}. As previously stated, typical ADM models leave this connection unexplained\footnote{See \cite{PhysRevD.89.063522} for an exception based on a common running of scales of the QCD and dark color group.}.
As a consequence of the lightness of the DM candidate, the DS cannot be charged under the SM given current collider bounds.
In such composite ADM models, the role of the new dynamics is not only to stabilize the DM candidate via residual symmetries, but to also open new avenues in the generation of the asymmetry, for example using dark sphalerons \cite{Blennow:2010qp}.
The rich spectrum of asymmetric composite dark matter models can be engineered to have peculiar signatures and cosmological histories \cite{Alves:2009nf,Detmold:2014qqa, Alves:2010dd}. In particular, the scarcity of anti-particles in the thermal bath can allow the formation of larger composite bound states like \emph{dark nuclei} and \emph{dark atoms}, leading to a very rich phenomenology \cite{PhysRevD.90.015023, Krnjaic:2014xza, Redi:2018muu, Hardy:2014mqa}. Besides, another hint to the asymmetric nature of the DM comes from the very value of the DM mass, $M_{\text{DM}}$. Indeed, ACDM models in their symmetric version predict $M_{\text{DM}}\simeq 100$ TeV in order to explain the observed DM abundance. The presence of an asymmetry significantly lowers this value, so that measuring a DM mass in the 50 TeV ballpark would clearly exclude the symmetric scenario. This, together with many different signatures pointing to a composite dark sector (detection of Gravitational Waves (GW) from the confinement phase transtion, characteristic direct detection signatures...), makes asymmetric ACDM models a phenomenologically distinct possibility.\\
Indeed, the goal of this paper is to start from ACDM models and try to extend them in the most minimal way to build an UV completion that can dynamically generate the asymmetry for the DM candidate, and deplete its symmetric component. The completion should also be technically natural and not spoil the accidental IR symmetries responsible for the DM stability.
Notice that ACDM models differ from the previous composite asymmetric models presented above: the dynamical scale is unrelated to the EW scale, since it is set by requiring the correct DM relic abundance, and its values are typically in the range $10 \div 100$ TeV. Moreover, their field content is not chiral under the EW group, making the use of EW sphaleron impossible.
Given the non-chiral nature of the dark quarks field content under dark color, also dark sphalerons are precluded.
Since one of the main features of ACDM models are collider signatures, the mass of a ACDM candidate cannot be at the GeV scale. This implies that the asymmetries of the visible and dark sector cannot be related using standard model building tools as done in light composite DM models \cite{PhysRevD.42.3344}. While there are mechanisms that allow a hierarchical asymmetry transfer \cite{Falkowski:2011xh}, they typically rely on some coincidence of the DM scale with the scales of other processes, making such mechanisms fine-tuned if applied to the ACDM scenario \cite{Buckley:2010ui}. The additional constraint of having confined dark quarks in the model does not allow the possibility to have a Higgsed phase for the dark color, that could induce a first order phase transition \cite{Shelton:2010ta, DUTTA2011364}.\\
Given the model-building limitations stemming from the defining features of ACDM models, we will not pursue the route of relating the visible and DM asymmetries. We focus only on the generation of the DM asymmetry, leaving the SM asymmetry to be generated in an independent, unspecified way. Despite this, we find that there is indeed the possibility of generating simultaneously the SM and DS asymmetries in one particular realization of our models. We will briefly present it without working out the details.\\
The structure of the paper is the following: in Section \ref{sec:review} we review the ideas behind ACDM models, and their specific field content.
In Section \ref{sec:making_acdm_asym} the necessary conditions needed to asymmetrize ACDM models are given, with a step-by-step description of the features of the resulting models.
In Section \ref{sec:benchmarkmodel} the basic building block of the asymmetrization procedure of ACDM models is given. In Section \ref{sec:buildingmodels} we present an asymmetric extension for each of the original \emph{golden class} ACDM models, that feature a stable, asymmetric DM candidate; the asymmetrization of such models is based on the benchmark implementation, or slight modifications of it, described in Section \ref{sec:benchmarkmodel}. In Section \ref{sec:othermodels} we complete the discussion by giving additional models that features an unstable DM candidate based on a slightly different implementation of the asymmetrization procedure. In Section \ref{sec:pheno}, we review the main phenomenological signature of Asymmetric ACDM models. In Section \ref{sec:conclusion} we summarize the results and discuss alternative possibilities to generate the asymmetry in ACDM models to be explored in future works.
\section{Review of Accidental Composite Dark Matter models}\label{sec:review}
The idea behind Accidental Composite DM models \cite{Antipin:2015xia} is to provide a DM candidate which is stable thanks to accidental symmetries in the Lagrangian, in a similar fashion to proton stability and baryon number conservation in QCD. The visible sector is thus enlarged with a Dark Sector (DS) made of new fermions $\Psi$, called \textit{dark quarks} (DCquark, DCq), charged under a new \textit{dark color} (DC) interaction, based on $\mathrm{SU}(N)_{\mathrm{DC}}$ or $\mathrm{SO}(N)_{\mathrm{DC}}$ gauge groups \footnote{Other confining gauge group like $\mathrm{Sp}(2N)$ or special group like $\mathrm{G}_2$ were not considered in the original paper. Here we do not aim at making an analogous classification of DM models for the missing groups, so that we stick to the groups mentioned in the main text.} that confines at a scale $\ldc$. The dark quarks are assumed to be in the fundamental representation of dark color, as well as vector-like representation under the SM. In particular, SM representations are taken to be "fragments" of the $\mathrm{SU}(5)$ GUT extension of the SM gauge group: the GUT framework motivates the choice of possible gauge representations for the DCquarks.\\
The DS renormalizable Lagrangian is simply given by:
\begin{equation}
\mathcal{L}_{\mathrm{DS}} = -\frac{1}{2}\Tr[\mathcal{G}_{\mathrm{D}}^{\mu\nu}\mathcal{G}_{\mathrm{D},\mu\nu}]+\overline{\Psi}_i(i\slashed{D}-m_{\Psi})\Psi_i+y_{ij}\overline{\Psi}_i\Psi_j H+\hc \;,
\end{equation}
where we have included a Dirac mass term $m_\Psi$ for the DCquarks, given their vector-like nature.
Below DC confinement, the dark quarks bind into \textit{dark hadrons}:

\begin{itemize}
    \item \textit{dark pions} (DC$\pi$), $m_{\mathrm{DC}\pi}^2 \approx m_\Psi \ldc$.
    \item \textit{dark baryons} (DCb), $\mtcb \approx \ndc \ldc$.
\end{itemize}

In $\mathrm{SU}(N)_{\mathrm{DC}}$ models, the lightest DCb is stabilized by an accidental $\udb$, the \textit{dark baryon number}, under which all dark quarks rotate with the same phase, and by an accidental $\mathbb{Z}_2$ in $\mathrm{SO}(N)_{\mathrm{DC}}$ models. The SM quantum numbers of the dark quarks must thus be chosen properly in order for this lightest DCb to be neutral and hence a good DM candidate. Charged \tcp are in general dangerous if protected by additional species symmetries, which must be broken either by Yukawas or by suitable higher dimensional operators. We focus on the so called \emph{golden class models} (GC), in which all the extra species symmetries responsible for \tcp stability are broken by Yukawa with the SM Higgs. Neutral \tcp instead are always unstable at the level of 5d operators with the Higgs generated at the Planck scale $M_{\mathrm{P}}= 1.22\times 10^{19}$ GeV, and are never good DM candidates. In Appendix \ref{sec:goldenclass} we list the original GC models. 
In Appendix \ref{sec:silverclass} we briefly comment on the possibility of applying our mechanism to \emph{silver class models}, that unlike GC models contain dangerous stable \tcp.\\
The cosmological evolution of ACDM models is rather simple and depends only on $\ldc$ and $m_{\Psi}$. If $\ldc\gg m_{\Psi}$, after DC confinement the DCb's go through a phase of non-perturbative annihilation, whose freeze-out determines the final DM abundance. Since in this regime both the DM mass and its annihilation cross-section are set solely by $\ldc$, the DM abundance turns out to be a function of this parameter alone. In the absence of any pre-existing dark baryon asymmetry, the observed $\Omega_{\mathrm{DM}}h^2\approx 0.119$ is reproduced for $\mtcb \approx 100$ TeV.\\
If $\ldc\ll m_{\Psi}$, instead, the cosmological evolution consists of two stages \cite{Mitridate:2017oky}: a first phase of perturbative annihilations among DCq's, which freezes-out around $T\approx m_{\Psi}/25$, followed, after DC confinement, by a phase of re-annihilation among DCbs. Indeed, in this regime the binding energy of DCbs is dominated by the Coulomb potential among the constituents rather than by confinement effects, so that the annihilation cross-section now is set by the Bohr radius $r_B\approx \left( \alpha_{\mathrm{DC}}m_{\Psi}\right)^{-1}$ of the bound state. Since $r_B\ll m_{\Psi}^{-1}$, the DCb-$\overline{\text{DCb}}$ annihilation cross-section is much larger than that among the constituent DCq (see \cite{Contino:2018crt} for a detailed discussion of this regime in composite DM models).
For simplicity, we will only analyze in the rest of the work the case $m_\Psi \ll \ldc$.
The presence of a pre-existing net dark baryon asymmetry alters in no way the different cosmological histories that we have outlined above: the requirement of annihilating the symmetric component only yields a different relation between the observed relic density $\Omega_{\mathrm{DM}}$ and $\ldc$ or $m_\Psi$.
\section{Making Accidental Composite DM Models Asymmetric}\label{sec:making_acdm_asym}
In this section we explore the different possibilities in order to make ACDM models asymmetric.
Our goal is to write the minimal UV completions of the models of \cite{Antipin:2015xia} that, at some UV scale $\luv$, can accommodate an asymmetry generation mechanism in the dark sector. Below $\luv$, we want to recover, at the EFT level, the original ACDM models. In other words, we want to build a UV completion for  ACDM models with an initial non-zero asymmetry at the cutoff $\luv$. Of course, an asymmetric DM model makes sense only if it is possible to distinguish the DM candidate from its antiparticle. This very basic requirement already makes all the $\mathrm{SO}(N)_{\mathrm{DC}}$ models classified in \cite{Antipin:2015xia} not suitable for asymmetrization: all the DM candidates in this case are of the form $(\overline{\Psi}_C\Psi_C)^n\Psi_R^m$, where $\Psi_C$ ($\Psi_R$) is a DCq in a complex (real) representation of the SM\footnote{In particular, such candidates do not carry any net species number.}. Therefore, we shall assume in the following that the DS is charged under a $\mathrm{SU}(N)_{\mathrm{DC}}$ gauge group.\\
We stress again that in this work we do not attempt to give a common explanation to the asymmetry of the dark sector and of the visible one.
One of the consequences of making DM asymmetric is to force a lower mass of the DCb $\mtcb$ (or equivalently the confinement scale $\ldc$) in order to satisfy cosmological constraints. Also, asymmetric dark matter models have different constraint coming from indirect detection, and in general have different phenomenology, which will be explored in section \ref{sec:pheno}.
In the following sections we briefly sketch the necessary ingredients needed to build a succesful asymmetric ACDM model, from the generation of the correct amount of asymmetry to the annihilation of the symmetric component.
\subsection{Sakharov conditions and DCb number}
In order to generate an asymmetry in the DS, the model must satisfy the three Sakharov conditions \cite{Sakharov:1967dj}:
\begin{enumerate}
    \item The presence of an out-of-equilibrium process.
    \item C and CP violation.
    \item Violation of the number species of the candidate to be asymmetrized.
\end{enumerate}
The last condition is particularly delicate: in $\mathrm{SU}(N)_{\mathrm{DC}}$ models, the very same symmetry that we need to break in order to generate the asymmetry is the $\udb$ responsible for accidental DM stability. Therefore in order to satisfy the Sakharov conditions, this symmetry must be broken at some scale, possibly associated to new extra fields, but at the same time it must be recovered at the EFT level: indeed, our goal is to keep intact the IR physics of original ACDM models. This is a further constraint on the UV completion we are looking for: it must not mediate a fast decay of the DCb in order to describe the observed relic density. This in the spirit of \cite{Perez:2013nra}, in which the authors engineered models of asymmetric dark matter with a gauged $\mathrm{U}(1)_{\mathrm{B-L}}$ that was spontaneously broken in the UV, while kept as a global (approximate) symmetry in the IR, stabilizing the non-composite DM candidate. Instead here we explicitly break a global, ungauged symmetry, and do not rely on spontaneous symmetry breaking, since it would break $\sun$ and forbid confinement.\\
In looking for minimal UV completions of asymmetric ACDM models, we restrict ourselves for simplicity to renormalizable UV Lagrangians.\\
\subsection{How to break DCb number}\label{sec:tcbbreaking}
Since ACDM models are vector-like, there is no analogue of the EW sphaleron \cite{KUZMIN198536} in the dark sector, unless we add a new gauge group under which DCquarks are chiral\footnote{This might be problematic because even in simple models, the additional copies of DCquarks due to the new gauge group might bring SM Landau poles below the GUT scale.}. For the same reason, the DCquarks do not couple to the EW sphalerons. This implies that there is no way to break $\udb$ via non-perturbative effects, and we must resort to perturbative terms in the Lagrangian.
If, by hypothesis, we restrict to renormalizable Lagrangians, adding only extra fermions will not allow to break in any way $\udb$. Indeed the only other terms compatible with $\sun$ invariance are bilinear in the dark fermions: the extra gauge terms and Yukawas between the Higgs and another DCquark, in which the fermions form a $\sun$ singlet. No $\sun$ invariant bilinear can break $\udb$, since the complex nature of the $\sun$ representations of DCquarks prevents real bilinears.
Consistently with our hypothesis of minimality, we enlarge the DS with the addition of a scalar field $\phi$. This scalar must fill a complex representation of $\sun$ in order to carry a non-trivial $\udb$ charge. Since we want to leave untouched the light spectrum of the original models, we take this scalar to be heavier than $\ldc$ and/or $m_\Psi$, so that it can later decay and disappear from present-day DM content. The only possible terms that we can add to the renormalizable interaction Lagrangian of the DS are:
\begin{equation}
\label{eq:udb_break}
\phi\Psi\Psi,\quad \phi\Psi\psi_{\mathrm{SM}}, \quad V(\phi,H) \;,
\end{equation}
where $V(\phi,H)$ is the scalar potential. Notice that we have not included terms of the form $\phi\overline{\Psi}\Psi$: they do not violate the $\mathrm{U}(1)$ associated to the $\Psi$ number. Moreover, if the scalar potential is to break $\udb$, it must contain terms of the form:
\begin{equation} 
\label{eq:scal_pot}
V(\phi,H) \supset \phi^3,~ \phi^4,~ \phi^3H^* \;.
\end{equation}
The use of such terms to break the dark baryon number has been explored previously in symmetric \cite{Cline:2016nab} and asymmetric \cite{Krnjaic:2014xza} (although non ACDM) contexts. Similar ideas can also be found in the study of baryon number-violating processes in the SM alone \cite{PhysRevD.87.075004,Helset:2021plg}.
Terms containing the real combination $\phi^2$, possible only for real $\phi$, are not viable as will be shown in Section \ref{sec:othermodels}.
Notice that we need at least two terms involving the dark scalar to break the global $\udb$
, since each operator in \eqref{eq:udb_break} is invariant under $\udb$ upon a different assignment of a dark baryon charge to $\phi$. If there is such a surviving $\udb$, after the scalars decay the conservation of this enlarged $\udb$ implies that no asymmetry can be generated in the DS sector, assuming it is stored in the single lightest DCb species. \\
This last condition leads to three different realizations of Asymmetric ACDM models, according to the pair of $\udb$ violating operators present in the interaction Lagrangian:
\begin{itemize}
    \item $\phi \Psi \Psi + V(\phi,H)$ (\emph{class 1} models): in Section \ref{sec:buildingmodels} we will present models in which, despite the breaking of $\udb$ at the renormalizable Lagrangian level, DM stability is guaranteed up to 5d operators included.
    \item $\phi \Psi \psi_{\mathrm{SM}}+ V(\phi,H)$ (\emph{class 2} models): in Section \ref{sec:othermodels} we will show that in this class of models the DM is unstable but long-lived.
    \item No $\udb$-violating scalar potential (\emph{class 3 models}). There are several possibilities to realize such models, depending on the nature of the two Yukawa portals: a "dark" Yukawa $\phi \Psi \Psi$ and a "mixed" Yukawa $\phi \Psi \psi_{\mathrm{SM}}$, two mixed Yukawa, two dark Yukawa. In Section \ref{sec:othermodels} we will show that for $\phi$ below the Planck scale, the effective operator obtained by integrating it out leads to fast decays of the lightest DCb.
\end{itemize}
Therefore, we can consistently generate an asymmetry in the DS without spoiling the stability of the DM candidate only if the Lagrangian contains both a Yukawa and a $\udb$-violating term in the scalar potential. In this case, since the interactions of $\phi$ with the dark quarks constrain the possible representations under $\sun$ to:
\begin{equation}
\Yvcentermath1
\phi\in \Yboxdim{12pt}
\yng(1)~,~ \yng(1,1)~,~ \yng(2) \;.
\end{equation}
the only $\sun$ gauge groups compatible with the terms in \eqref{eq:scal_pot} are those with $\ndc=3, 4, 6, 8$.
We remark that the role of the scalar is not to spontaneously break the dark color: the Higgsed phase inevitably makes the DM unconfined in the IR regime, and therefore with dangerous SM charges, excluding some highly non-trivial scenarios in the spirit of \cite{Ramsey_Musolf_2018} in which the dark color symmetry is restored at low temperatures.

\subsection{Generating the asymmetry}
In order to build a successful asymmetric dark matter model, we must specify a proper out-of-equilibrium process to satisfy the Sakharov conditions. 
The easiest way to accomplish this is to consider heavy scalars (for example $M_\phi \gtrsim 10^{15}\;\mathrm{GeV}\sim M_{\mathrm{GUT}}$ as shown in Section \ref{sec:asym_comp}) that decay as soon as the temperature of the plasma drops below its mass, like in the original GUT baryogenesis scenarios \cite{PhysRevD.20.2484}.
A necessary condition for the mechanism to work is that the scalar must have access to multiple decay channels with different dark baryon number in the final states.
In viable models, a single heavy $\phi$ can decay to channels with different $\udb$ by inserting the pair of $\udb$-breaking operators chosen (for example the Dark Yukawa and the $\phi ^3$ term in the models of Section \ref{sec:buildingmodels}). However  in this case the asymmetery turns out to be suppressed as we explain in Section \ref{quick_est}. This is in part due to the fact that all the DCquarks have the same $\udb$ charge by hypothesis, and therefore the different decay channels must involve different number of DCquarks, leading to processes with many insertions of the couplings.
A simple solution to avoid the large suppression of the interference term, is to consider a second flavor of $\phi$. For example, in models of Section \ref{sec:buildingmodels}, if the scalars are not degenerate, the decay
\begin{equation}
\phi_H \rightarrow \phi_L^\dagger \Psi \Psi    
\end{equation}
is allowed.
As we will see, this decay violates the dark baryon number (if assigned to be conserved in the two-body decay), and therefore there are two channels with different dark baryon number for the heavy $\phi$ decay. Subsequent decays of the lighter $\phi$ do not generate further asymmetries: they simply transfer to the other DCquarks their dark baryon number (up to the negligible asymmetry due to $\phi$ decays in more than 2 DCquarks).
It's important to notice that in order for this scenario to work, the inverse decay processes must be suppressed, otherwise they will wash out any generated asymmetry. As we will show in Section \ref{sec:asym_comp}, this translates into a bound on the mass of the heavy scalar, at fixed coupling, called the \emph{weak wash-out condition}. For dimensionless Yukawa couplings $ y\sim \mathcal{O}(0.1)$, this amounts to $M_\phi\gtrsim 10^{15}$ GeV. Besides, it is possible to show that these couplings get (non-multiplicatively) renormalized only at two-loop order. Hence, technical naturalness sets a lower bound on the Yukawas that can be estimated as $y\geq (16\pi^2)^{-2} \sim 10^{-4}$. This implies that the smallest scalar mass compatible with the weak wash-out condition and naturalness is $M_{\phi}\sim 10^{10}$ GeV.
\subsection{Annihilating the symmetric part}\label{sec:annihilation}
\begin{figure}[htp!]
    \centering
    \includegraphics[scale=0.6]{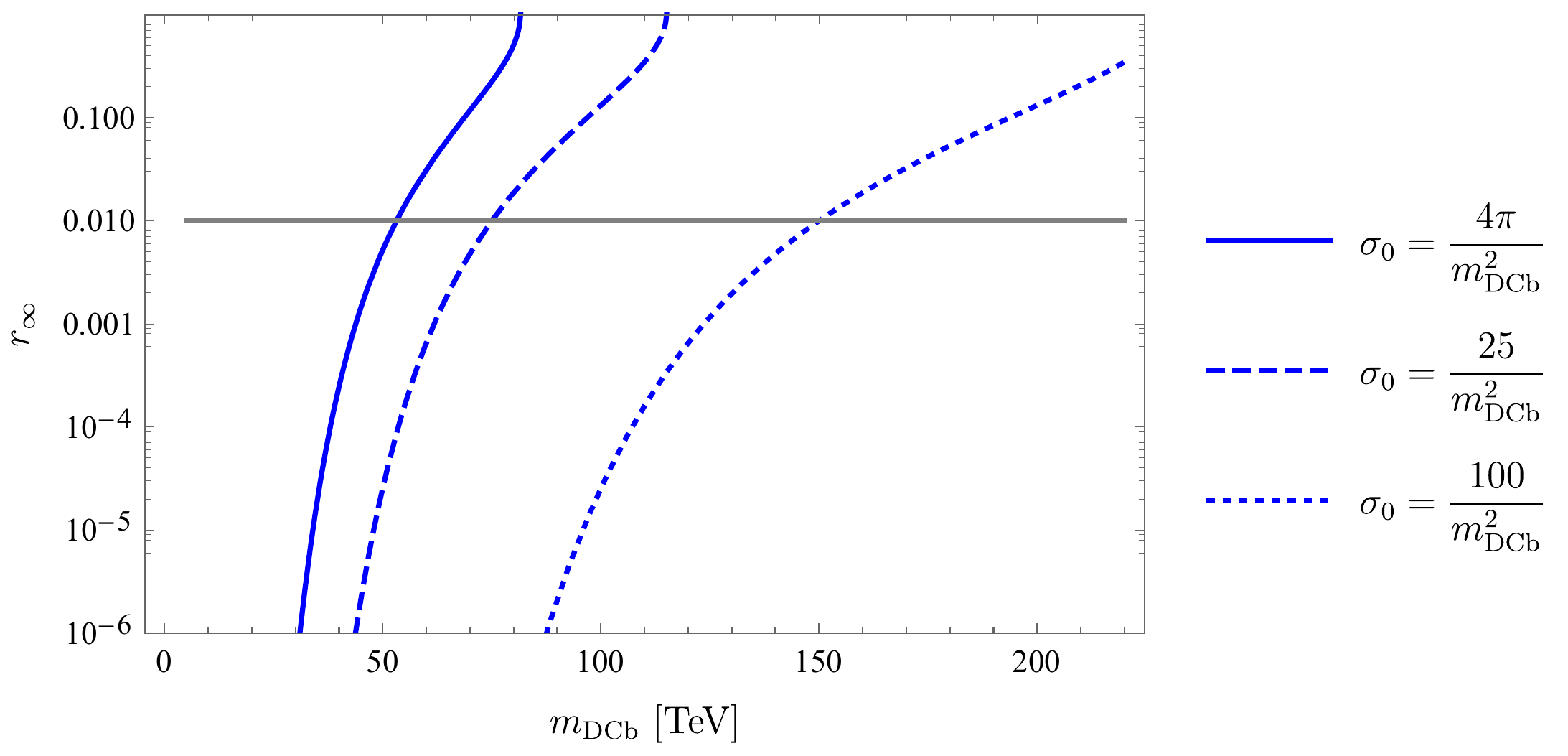}
    \caption{$r_\infty$ as a function of $\mtcb$ assuming only non-perturbative annihilations. The solid line assumes a cross-section that saturates the perturbative unitarity limit. The dotted line corresponds to the annihilation cross-section obtained in QCD, which is 10 times larger. Given the large uncertainty, we shall use an intermediate value that reproduces the typical DM mass in ACDM models, $\mtcb\approx 100$ TeV, as shown by the dashed line. Finally, the gray line represents the reference value $r_\infty=0.01$.}
    \label{fig:rinf}
\end{figure}

Independently on the mechanism responsible for the generation of the asymmetry, in all these models it's possible on general grounds to relate the dynamical scale to the amount of asymmetry necessary to reproduce the correct DM abundance, requiring at the same time that the non-perturbative annihilations are sufficient to deplete the abundance of the symmetric DM component. In fact, if $X$ is the DM and $\overline{X}$ its antiparticle and if we define

\begin{equation}
    r\equiv \frac{n(\overline{X})}{n(X)} \;,
\end{equation}

then solving the Boltzmann equation in the presence of an asymmetry gives the following relation between $r$ evaluated at late times, $r_\infty$, and the thermally averaged annihilation cross-section $\langle \sigma v\rangle\equiv\sigma_0$ \cite{Petraki:2013wwa},\cite{Graesser:2011wi}:

\begin{equation}\label{eq:anni_sym}
    r_\infty \approx \exp\left[-2\left(\frac{\sigma_0}{\sigma_{\mathrm{WIMP}}}\right)\left(\frac{1-r_\infty}{1+r_\infty}\right)\right] \;,
\end{equation}

where $\sigma_0$ can be estimated as $\sigma_0 \approx \frac{25}{\mtcb^2}$ and $\sigma_{\mathrm{WIMP}}\equiv\left(\frac{1}{23~\text{TeV}}\right)^2$. In Figure \ref{fig:rinf} we show how $r_\infty$ depends on $\mtcb$. As we can see, we need $\mtcb \lesssim 50\div 75$ TeV in order to have $r_\infty \lesssim 0.01$. Equivalently, discovering that the DM is composite and with mass in such range explicitly points to an asymmetric DM content.

At this point we can relate the dynamical scale to the asymmetry $\eta_{\mathrm{DM}}$  \cite{Petraki:2013wwa}:

\begin{equation}
\label{etads}
    \eta_{\mathrm{DM}}\approx 5 \frac{m_p}{\mtcb} \eta_b~ \frac{1-r_\infty}{1+r_\infty} \approx 2.2 \times 10^{-14}\left(\frac{23~\text{TeV}}{\mtcb}\right) \;,
\end{equation}
where $m_p\approx 1$ GeV is the proton mass.
Notice that, unlike previous models such as in \cite{Krnjaic:2014xza}, we do not introduce any new gauged $\mathrm{U}(1)$ to deplete the symmetric component, and we only have a single DM candidate, not two oppositely charged species sharing the asymmetry.
\subsection{Brief cosmological history}
Once the asymmetry has been generated from the decay of the heavy $\phi$, the IR theory is essentially a standard ACDM model, with a non-zero asymmetry as initial condition for the (approximately) conserved $\udb$ charges.
Above $\ldc$, annihilation between DCquarks and their antiparticle are fast, and the two species are kept in chemical equilibrium.
Below $\ldc$, DCquarks will form DCbs and \tcp. In the GC model under scrutiny, the latter will decay into SM particles via Yukawa with the SM Higgs, or through a 5d operator generated at $M_{\mathrm{P}}$. Since in GC models all species symmetries are broken by the Yukawas\footnote{An exception to this is the $V \oplus N$ model, on which we will comment in Section \ref{sec:phi3}.}, the various DCbs will undergo fast decays into the lightest one via the species number-breaking Yukawas with the Higgs  once the masses are split due to SM gauge interactions\footnote{The mass difference can also be induced by different DCquarks bare masses.}. Since all these interactions conserve $\udb$, the asymmetry will be stored and conserved in the lightest DCb containing the constituents that possess an initial asymmetry. The condition on $\mtcb$ (eq. \ref{eq:anni_sym}) guarantees that there is no symmetric part left.
In some asymmetric ACDM models, the possibility to radiate light SM states allows the DCbs to form \emph{dark nuclei} \cite{Redi:2018muu, Mahbubani:2019pij, Mahbubani:2020knq}, for temperatures below the typical dark nuclear binding energy $E_B$, with distinct phenomenological signatures that will be explored in Section \ref{sec:pheno}.
In some models the presence of the scalar in the UV Lagrangian (or equivalently of the specific asymmetry generation mechanism) is reflected in the IR with a Majorana mass term for the DCb, that can cause oscillations between the DCb and its antiparticle. Such effect will be discussed in Sections \ref{sec:tcboscillation} and \ref{sec:pheno_osc}.

\section{Benchmark model}\label{sec:benchmarkmodel}
In this section we will describe a simple model to show the mechanism of asymmetry generation and ensure that we can naturally obtain the correct amount of asymmetry. This benchmark model can be easily adjusted to fit in every ACDM model, as will be shown in Sections \ref{sec:buildingmodels}, \ref{sec:othermodels}.
In particular, we consider as a benchmark case $\Psi=N$, $\ndc=3$, as shown in table \ref{tab:benchmark_fields}.
\begin{table}[!h]
    \centering
    \begin{tabular}{|c|c|c|c|}
        \hline
        Field & $\sut$ & $\left(\mathrm{SU}(3)_c,\mathrm{SU}(2)_L \right)_Y$ & $\udb$ ($D$)\\
        \hline
        $N$ & $3$ & $(1,1)_0$ & $1$\\
        \hline
        $\phi$ & $\bar{6}$ (sym) & $(1,1)_0$ & $-2$ \\
        \hline
    \end{tabular}
    \caption{Benchmark model field content.}
    \label{tab:benchmark_fields}
\end{table}\\
We will consider the following Lagrangian:
\begin{equation}\label{eq:benchmark_lagrangian}
    \mathcal{L}=\mathcal{L}_{\mathrm{kin}}+y\phi_{ij}N^iN^j+\lambda M_\phi \epsilon^{ijk} \epsilon^{i'j'k'}\phi_{ii'}\phi_{jj'}\phi_{kk'}+ \lambda_4 (\phi^\dagger \phi)^2 \;,
\end{equation}
where $\epsilon$ stands for the totally antisymmetric Levi-Civita tensor of $\sut$.
In equation \ref{eq:benchmark_lagrangian} $\mathcal{L}_{\mathrm{kin}}$ contains the kinetic and mass terms for $\phi$ and $N$, $M_\phi$ is the mass of $\phi$, which is taken to be near the cutoff scale, and $\lambda$ is a dimensionless parameter.
Notice that in presence of SM fields, the only additional term allowed by gauge invariance and renormalizability to the ones in equation \ref{eq:benchmark_lagrangian} is $\phi^\dagger \phi H^\dagger H$, which plays no role in $\udb$ asymmetry generation. 
A similar model can be found in \cite{Krnjaic:2014xza}, although employed in a different context.
We remark that another possible benchmark model, in which the computations are essentially the same, is obtained by replacing $N$ with the $\mathrm{SU}(2)_L$ triplet $V$ DCquark: indeed the Yukawa $\phi VV$ can still exist if $\phi$ is a SM singlet by contracting the $\mathrm{SU}(2)_L$ indices of the $V$s together. \\
First, we will show a model with two flavors of $\phi$, in which the asymmetry is generated via the out-of-equilibrium decay of the heaviest of the flavor of $\phi$. This implementation leads to the correct amount of asymmetry, and will be employed in Sections \ref{sec:buildingmodels},\ref{sec:othermodels}.
Then, for completeness we will give a brief overview of the problems encountered when trying to generate the asymmetry using a single flavor of scalar. In this case, we try to generate the asymmetry both via decay and via the thermal freeze-out of the number-changing processes involving $\phi$. We give heuristic arguments to show that this process does not lead to a large enough asymmetry.\\
As a final comment, we point out that the values of the couplings appearing in the potential should be such that the vacuum does not break $\sun$, without tuning excessively the parameters (if tuning is allowed, the potential can always be made positive by taking for example $\lambda$ close to $0$). For $\sut$ as gauge group and a single scalar flavor it has been shown in \cite{Bai:2017zhj} that this is indeed possible by taking the couplings in a natural region in parameter space. With additional flavors, it should still be possible to realize this scenario without fine tuning by observing that a trilinear term in the potential can always be bounded between the sum of quadratic and quartic terms in the potential, up to $\mathcal{O}(1)$ factors. In later sections, we will always assume that it's possible to not break $\sun$ by appropriately choosing the parameters, and that this procedure does not introduce unwanted fine tuning.
\subsection{Two flavors of $\phi$: a possibility}\label{sec:2phi}
In this section we study the possibility to generate the asymmetry by considering the benchmark model with two flavors of $\phi$ (which we remind is a total SM singlet and is in the conjugate 2-symmetric of $\mathrm{SU}(3)_{\mathrm{DC}}$). The Yukawa coupling $y$ and the trilinear coupling $\lambda$ now carry also scalar flavor indices. Notice that the addition of the second flavor upgrades the $\mathrm{U}(1)$ related to $\phi$ rephasing to an $\mathrm{SU}(2)$. However, this symmetry is broken to the diagonal rephasing $\mathrm{U}(1)$ by different mass terms, and by the couplings $y_i, \lambda_{ijk}$.
The easiest way to generate the asymmetry is to mimick GUT baryogenesis \cite{PhysRevD.20.2484}. In this framework, the asymmetry is generated via the out-of-equilibrium decay of a heavy scalar.
In order to do so we take a scalar $\phi_H$ heavier than the second $\phi_L$: $M_H\gtrsim M_L$. Avoiding the hierarchy between the two flavors of $\phi$ keeps the model natural. For the sake of showing that the mechanism indeed works, we pick a specific region of the parameter space. In particular, we take the masses of the scalars close:  $M_L<M_H\lesssim 2M_L$. In this way we avoid the possibility of $\phi_H$ decaying into a pair of light scalars $\phi^\dagger_L\phi_L^\dagger$  (however this is not mandatory and does not affect the discussion significantly).
We stress that in this mechanism we do not need quasi-degenerate state, and the previous request is not a fine-tuning of the parameters. Both masses are taken to be much heavier than $\max \left(\ldc,m_N\right)$, so that the IR spectrum of DCquark bound states is untouched. The complete interaction Lagrangian therefore reads:
\begin{equation}
\label{eq:int_lag}
\begin{split}
    \mathcal{L}_I =~& \frac{y_H}{2}\phi_H  \overline{N}^c N+\frac{y_L}{2}\phi_L \overline{N}^c N +\frac{\lambda_{HHH}}{6}M_H\phi_H^3+\frac{\lambda_{HHL}}{2}M_H\phi_H^2\phi_L\\
    &+\frac{\lambda_{HLL}}{2}M_H\phi_H\phi_L^2+\frac{\lambda_{LLL}}{6}M_H\phi_L^3+\hc \;,
\end{split}
\end{equation}
where we have taken the Yukawa couplings $y_{H,L}$ to be the same for the right-handed and left-handed components of the Dirac DCquark $N$. The couplings have been divided by numerical factors accounting for the symmetry factor in Feynman diagrams.
In the model the heavy scalar has access to two decay channels with different dark baryon number $D$:
\begin{itemize}
    \item $\phi_H \rightarrow \bar{N}\bar{N}$;~ $\Delta D=0$
    \item $\phi_H \rightarrow \phi^\dagger_L N N$;~ $\Delta D = 6$
\end{itemize}
where $\Delta D$ stands for the difference of $D$ charge between final and initial states.
In principle $\phi_H$ can undergo a $4$-body decay into $NNNN$, by simply attaching $NN$ to the $\phi_L$ leg in the 3-body decay graph. This decay has the same $\Delta D$ of the 3-body decay, and can give a contribution to the asymmetry generation. However this decay is suppressed with respect to the 3-body decay by an additional coupling insertion and propagator suppression: it reduces to the 3-body decay only if the internal leg can go on-shell, which is true when it's $\phi_L$. Therefore we will neglect this extra contribution.\\
The presence of the multiple decay channels with different $\Delta D$ is what allows the asymmetry generation.
Indeed, let $\Gamma , \Gamma_2$ be the total decay width, and the decay width for the 2-body decay $\phi_H \rightarrow \bar{N}\bar{N}$ channel  (with final $D=-2$) respectively. Neglecting further decay channels with respect to the two mentioned above, we can approximate the decay width in the $\phi^\dagger _L N N$ channel to be $\Gamma-\Gamma_2$. The decays of  $\phi_H$ will produce the following contribution to the total $D$ of the universe, accordingly to how the $\phi$ decays gets distributed in the two channels:
\begin{equation}\label{eq:phidecays}
    \frac{\Gamma_2 (-2D_N-D_\phi) + (\Gamma-\Gamma_2) (-2D_\phi+2D_N)}{\Gamma}\;.
\end{equation}
By calling $\bar{\Gamma}_2$ the decay width in $NN$ of $\phi_H^\dagger$, we get that the decays of $\phi^\dagger_H$ contribute as follows to the total $D$:
\begin{equation}\label{eq:phidaggerdecays}
    \frac{\bar{\Gamma}_2 (2D_N+D_\phi) + (\Gamma-\bar{\Gamma}_2) (2D_\phi-2D_N)}{\Gamma} \;,
\end{equation}
where by CPT invariance the total decay width of the particle and antiparticle are the same.
By putting together equations \ref{eq:phidecays}, \ref{eq:phidaggerdecays}, and assuming an equal initial abundance for $\phi_H$, $\phi_H^\dagger$, we get that the asymmetry in dark baryon number generated after the decay of $\phi_H$ is:
\begin{equation}
    \eta_{\mathrm{DM}}=\frac{n_{\phi_H}}{s}\frac{1}{\Gamma}\left(\Gamma_2 - \bar{\Gamma}_2  \right)\left( -4D_N +D_\phi  \right) \equiv \frac{n_{\phi_H}}{s} \epsilon \left( -4D_N +D_\phi  \right) \;.
\end{equation}
So after all the $\phi_H, \phi_H^\dagger$ are decayed, even if their initial abundances were equal, a net dark baryonic asymmetry is created\footnote{The presence of additional DCquarks does not spoil this argument since each decay channel has the same dark baryon number. This is because all the DCquarks are in the same representation of dark color and the dark baryon number can be assigned to be the $N$-ality of this representation}.
The asymmetry that is generated by the decay of the heaviest scalar is now split between the lighter scalar $\phi_L$ and the dark fermion $N$. If we can find a region in the parameter space in which we can neglect the $2\rightarrow 2$ processes involving these two lighter species, the $\phi_L$ again undergoes an out-of-equilibrium decay. The scalar $\phi_L$ decays dominantly in $\bar{N} \bar{N}$, but it can also decay in $NNNN$. Thus in principle an asymmetry can be generated also at this step of the decay chain. However, the asymmetry parameter $\epsilon_L$ associated to the decay of $\phi_L$ is negligible with respect to the asymmetry $\epsilon$ of the heavy scalar decay: it's suppressed by an additional $y^2$ and by an additional propagator suppression. Therefore the asymmetry generated by $\phi_L$ decays can be neglected when computing the total dark baryon asymmetry, and such processes will simply transfer the net asymmetry generated by the decay of the heavy scalar to the DCquarks.
After the decay of all the $\phi_L,\phi_L^\dagger$ no additional asymmetry will be generated, since there are no further active $\udb$-violating processes.\\
In order for this mechanism to be successful, we have to ensure two conditons: 
\begin{itemize} 
\item the presence of a complex phase in the decay amplitudes. It must come from both the phase in the couplings, and from the imaginary part of the graph associated to some particles going on shell in the propagators. The former is realized in our model by the presence of three phase-invariant, complex coupling combinations, as shown in Section \ref{sec:anotherestimate}: there are six couplings and 3 fields that can be rephased. 
The latter condition is accomplished considering fermionic bubbles inserted in the external scalar legs, or box diagrams with internal fermionic lines. Given that the mass of the scalars are both heavier than $2m_N$, the momentum circulating in such loops can make the virtual $N$s go on-shell, giving an imaginary contribution to the integral. Bubbles can also be inserted in the internal propagators, but via a direct computation it can be shown that such amplitudes do not contribute to the asymmetry.
\item We have to make sure that there is a region in the parameter space in which we can neglect washout processes coming from inverse decay and $2\rightarrow 2$ scatterings. As shown in Section \ref{sec:anotherestimate}, by taking the two scalars heavy enough the inverse decays can be neglected: when the temperature of the bath drops below the mass of the heaviest available scalar the inverse decay process are kinematically blocked. The same argument can be applied to the scattering processes in the very weak washout regime, in which the decays are always the last processes to fall out thermal equilibrium. This requires $\Gamma_{\mathrm{D}}/\hub(M_\phi) \ll 1$, which sets a condition on the mass of the scalar and the coupling entering the decay process (mostly the two Yukawas $y_i$).
\end{itemize}
Taking the scalars heavy has the side effect of guaranteeing that the running of SM and dark color couplings is not modified below $M_\phi$, avoiding the risk of Landau Poles at low energies.
 We stress that the freedom to take the scalars heavy comes with a price: the scalar sector will be hard to test. However, as will be shown in section \ref{sec:pheno_osc}, its presence gives rise to IR phenomena like DM-$\overline{\mathrm{DM}}$ oscillations that can be in principle testable.
\subsubsection{Estimating the conditions}\label{sec:anotherestimate}
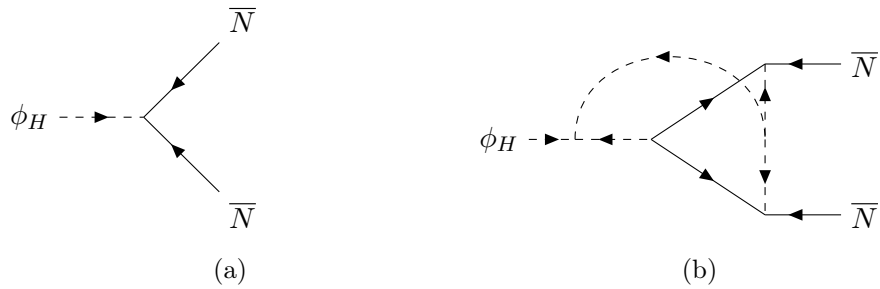
\begin{figure}[H]
\centering
    \begin{subfigure}[b]{0.4\textwidth}
    \begin{tikzpicture}[/tikzfeynman/small]
        \begin{feynman}
        \vertex (i){$\phi_H$};
        \vertex [right = 1.5cm of i] (v1);
        \vertex [above right = 1.4cm of v1] (f1){$\overline{N}$};
        \vertex [below right = 1.4cm of v1] (f2){$\overline{N}$};   
        \diagram*[small]{(i) -- [charged scalar](v1) -- [anti fermion](f1),
        (v1) -- [anti fermion](f2)};
        \end{feynman}
    \end{tikzpicture}
    \caption{}
    \label{fig:tree_1to2}
    \end{subfigure}
    \begin{subfigure}[b]{0.4\textwidth}
    \begin{tikzpicture}[/tikzfeynman/small]
        \begin{feynman}
        \vertex (i){$\phi_H$};
        \vertex [right = 1cm of i] (v1);
        \vertex [right = 1cm of v1] (l1);
        \vertex [right = 1.5cm of l1] (l3);
        \vertex [above = 1cm of l3] (l2);
        \vertex [below = 1cm of l3] (l4);
        \vertex [right = 1cm of l2] (f1){$\overline{N}$};
        \vertex [right = 1cm of l4] (f2){$\overline{N}$};
        \diagram*[small]{(i) -- [charged scalar](v1) -- [anti charged scalar](l1) -- [fermion](l2) -- [anti fermion](f1),
        (l2) -- [anti charged scalar](l3) -- [charged scalar](l4) -- [anti fermion](f2),
        (v1) -- [anti charged scalar, half left](l3),
        (l1) -- [fermion](l4)};
        \end{feynman}
    \end{tikzpicture}
    \caption{}
    \label{fig:loop1_1}
    \end{subfigure}
    \caption{Two-body decay of $\phi_H$ contributing to the generation of the asymmetry in $N$. The first loop diagrams appear only at two loops and an example is shown on the right. The arrows indicate the particle number flow.}
    \label{fig:twobody}
\end{figure}

\begin{figure}[H]
\centering
    \begin{subfigure}[b]{0.4\textwidth}
    \begin{tikzpicture}[/tikzfeynman/small]
        \begin{feynman}
        \vertex (i){$\phi_H$};
        \vertex [right = 1.5cm of i] (v1);
        \vertex [above right = 1cm of v1] (f1){$\phi\dg_L$};
        \vertex [below right = 1.2cm of v1] (v2);
        \vertex [above right = 1cm of v2] (f2){$N$};
        \vertex [below right = 1cm of v2] (f3){$N$};        
        \diagram*[small]{(i) -- [charged scalar](v1) -- [anti charged scalar](v2) -- [fermion](f2),(v2) -- [fermion](f3),
        (v1) -- [anti charged scalar](f1)};
        \end{feynman}
    \end{tikzpicture}
    \caption{}
    \label{fig:tree_1}
    \end{subfigure}
    \begin{subfigure}[b]{0.4\textwidth}
    \begin{tikzpicture}[/tikzfeynman/small]
        \begin{feynman}
        \vertex (i){$\phi_H$};
        \vertex [right = 1cm of i] (w1);
        \vertex [right = 1.cm of w1] (w2);
        \vertex [right = 0.75cm of w2] (v1);
        \vertex [above right = 1cm of v1] (f1){$\phi\dg_L$};
        \vertex [below right = 1.2cm of v1] (v2);
        \vertex [above right = 1cm of v2] (f2){$N$};
        \vertex [below right = 1cm of v2] (f3){$N$};        
        \diagram*[small]{(i) -- [charged scalar](w1) -- [anti fermion, half left](w2) -- [charged scalar](v1) -- [anti charged scalar](v2) -- [fermion](f2),(v2) -- [fermion](f3),
        (v1) -- [anti charged scalar](f1),
        (w1) -- [anti fermion, half right](w2)};
        \end{feynman}
    \end{tikzpicture}
    \caption{}
    \label{fig:loop_1}
    \end{subfigure}\\
    \begin{subfigure}[b]{0.4\textwidth}
    \begin{tikzpicture}[/tikzfeynman/small]
        \begin{feynman}
        \vertex (i){$\phi_H$};
        \vertex [right = 1.5cm of i] (v1);
        \vertex [above right = 1cm of v1] (w1);
        \vertex [above right = 1cm of w1] (w2);
        \vertex [above right = 1cm of w2] (f1){$\phi\dg_L$};
        \vertex [below right = 1.2cm of v1] (v2);
        \vertex [above right = 1cm of v2] (f2){$N$};
        \vertex [below right = 1cm of v2] (f3){$N$};        
        \diagram*[small]{(i) -- [charged scalar](v1) -- [anti charged scalar](v2) -- [fermion](f2),(v2) -- [fermion](f3),
        (v1) -- [anti charged scalar](w1) -- [fermion, half left](w2) -- [anti charged scalar](f1),
        (w1) -- [fermion, half right](w2)};
        \end{feynman}
    \end{tikzpicture}
    \caption{}
    \label{fig:loop_2}
    \end{subfigure}
    \begin{subfigure}[b]{0.4\textwidth}
    \begin{tikzpicture}[/tikzfeynman/small]
        \begin{feynman}
        \vertex (i){$\phi_H$};
        \vertex [right = 1.5cm of i] (v1);
        \vertex [above right = 1cm of v1] (f1){$\phi\dg_L$};
        \vertex [below right = 0.6cm of v1] (w1);        
        \vertex [below right = 1.cm of w1] (w2);
        \vertex [below right = 0.6cm of w2] (v2);    
        \vertex [above right = 1cm of v2] (f2){$N$};
        \vertex [below right = 1cm of v2] (f3){$N$};        
        \diagram*[small]{(i) -- [charged scalar](v1) -- [anti charged scalar](w1)-- [fermion, half left](w2)-- [anti charged scalar](v2) -- [fermion](f2),
        (v2) -- [fermion](f3),
        (v1) -- [anti charged scalar](f1),
        (w1) -- [fermion, half right](w2)};
        \end{feynman}
    \end{tikzpicture}
    \caption{}
    \label{fig:loop_3}
    \end{subfigure}\\
    \begin{subfigure}[b]{0.4\textwidth}
    \begin{tikzpicture}[/tikzfeynman/small]
        \begin{feynman}
        \vertex (i){$\phi_H$};
        \vertex [right = 1.5cm of i] (v1);
        \vertex [above right = 1cm of v1] (w1);
        \vertex [below right = 1cm of v1] (w2);
        \vertex [right = 1.41cm of v1] (w3);
        \vertex [right = 1cm of w1] (f1){$N$};
        \vertex [right = 1cm of w2] (f2){$N$};
        \vertex [right = 1cm of w3] (f3){$\phi\dg_L$};
        \diagram*[small]{(i) -- [charged scalar](v1) -- [anti fermion](w1) -- [charged scalar](w3) -- [anti charged scalar](w2) -- [fermion](v1),
        (w3) -- [anti charged scalar](f3),
        (w1) -- [fermion](f1),
        (w2) -- [fermion](f2)};
        \end{feynman}
    \end{tikzpicture}
    \caption{}
    \label{fig:loop_4}
    \end{subfigure}
    \begin{subfigure}[b]{0.4\textwidth}
    \begin{tikzpicture}[/tikzfeynman/small]
        \begin{feynman}
        \vertex (i){$\phi_H$};
        \vertex [right = 1.5cm of i] (v1);
        \vertex [above right = 1cm of v1] (w1);
        \vertex [below right = 1cm of v1] (w2);
        \vertex [right = 1.41cm of v1] (w3);
        \vertex [right = 1cm of w1] (f1){$N$};
        \vertex [right = 1cm of w2] (f2){$N$};
        \vertex [right = 1cm of w3] (f3){$\phi\dg_L$};
        \diagram*[small]{(i) -- [charged scalar](v1) -- [anti charged scalar](w1) -- [fermion](w3) -- [anti fermion](w2) -- [charged scalar](v1),
        (w3) -- [anti charged scalar](f3),
        (w1) -- [fermion](f1),
        (w2) -- [fermion](f2)};
        \end{feynman}
    \end{tikzpicture}
    \caption{}
    \label{fig:loop_5}
    \end{subfigure}
    \caption{Diagrams representing the three-body decays of $\phi_H$. The different flavors of scalars flowing in the internal lines allows for a non-zero imaginary part in their interference term. It is possible to show that no contribution comes from diagram \ref{fig:loop_3}. The arrows indicate the particle number flow.}
    \label{fig:threebody}
\end{figure}
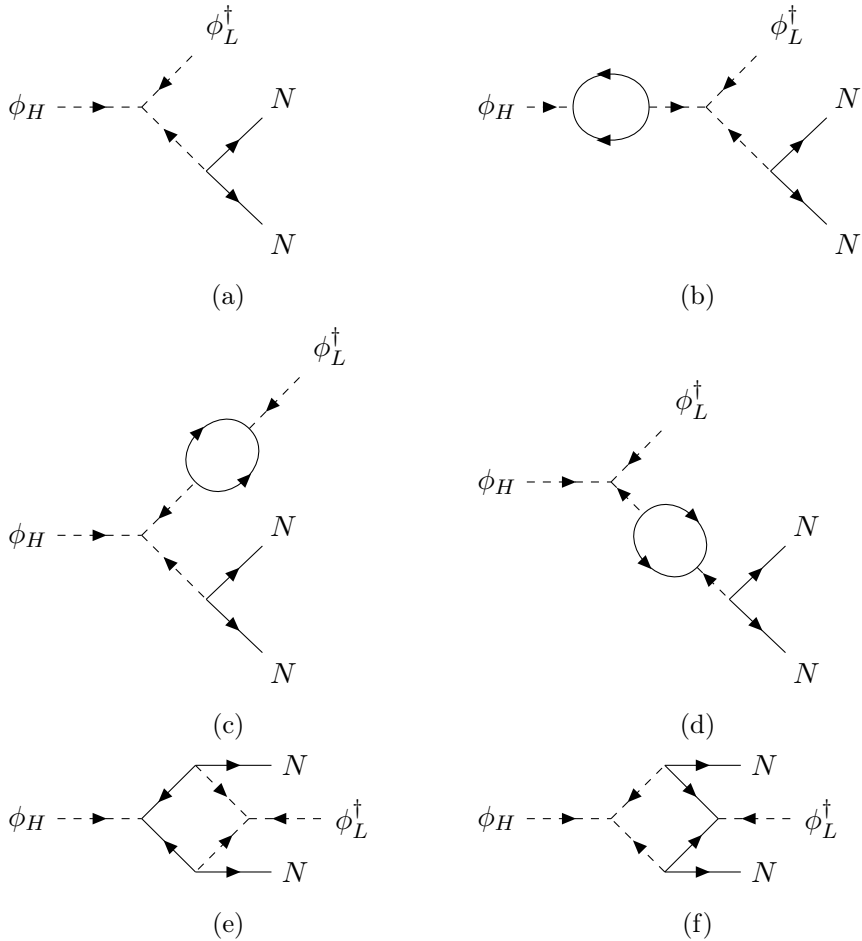

In this section we will study in detail the two conditions mentioned in Section \ref{sec:2phi}.
First we check if we can generate an interference in one of the decay channel to get a net CPV between the decays of $\phi_H, \phi_H^\dagger$.
In the two-body decays, the asymmetry is generated by the interference between the tree level diagram and a two-loop one as shown in Figure \ref{fig:twobody}.
In the three-body decays in Figure \ref{fig:threebody}, the asymmetry is generated by the interference among the tree level diagrams represented in Figure \ref{fig:tree_1}, that are identical up to the different virtual scalars circulating in the internal line, and the 1-loop diagrams in Figures \ref{fig:loop_1}-\ref{fig:loop_2}. Indeed, a closer look at the diagram \ref{fig:loop_3} shows that it does not contribute to the asymmetry.\\
It's interesting to check that the two asymmetries are related, as implied by CPT invariance and unitarity \cite{Kolb:1979qa}: indeed the three-body decay can be obtained by properly cutting the two-loops two-body decay.
A quick parametric estimate tells that the asymmetry factor is expected to be proportional to $\epsilon\sim \lambda^2 y^2/(16\pi^2)^2$. \\

The remaining question is whether we can work in a regime in which we can neglect the washout processes.

Following the discussion in \cite{Kolb:1990vq}, the strength of the wash-out processes is determined by the parameter:

\begin{equation}
\label{eq:weakwashout}
    K \equiv \left.\left(\frac{\Gamma_{\mathrm{D}}}{2\hub}\right)\right|_{T=M_H} \;,
\end{equation}
where $\Gamma_{\mathrm{D}}$ stands for a generic average decay rate of the scalars and $\hub$ for the Hubble parameter.
We are interested in $K\ll 1$, the so-called weak wash-out regime. In this regime, once that $T$ drops below $M_H$, decays are not efficient and the number of $\phi_H$ gets no exponential suppression, so that $n_{\phi_H}\propto n_\gamma$. This guarantees the required departure from equilibrium necessary to produce a net asymmetry. We are also implicitly assuming that $\phi_H$ was in equilibrium at some large temperature and that interactions with the thermal bath (\emph{e.g.} annihilations into dark gluons) are not efficient below $T=M_H$; this holds for masses larger than $\sim 10^{15}$ GeV.  Moreover, under our assumptions, it is possible to relate the two- and three-body decay rates $\Gamma$ to the rates of their inverse processes $\Gamma_{\mathrm{ID}}$ and show that they negligible. Indeed, these inverse rates are given by:

\begin{equation}
\label{eq:inv_dec}
\left\{
    \begin{split}
    \Gamma_{\mathrm{ID2}} & = \frac{n_{\phi_H}^{eq}}{n_N^{eq}}\Gamma_{\mathrm{D2}}\approx \left(\frac{M_H}{T}\right)^{\frac{3}{2}}e^{-\frac{M_H}{T}}\Gamma_{\mathrm{D2}}\\
    \Gamma_{\mathrm{ID3}} & = \frac{n_{\phi_H}^{eq}}{n_{\phi_L}^{eq}}\Gamma_{\mathrm{D3}} \;,
    \end{split}
\right.
\end{equation}
where we have defined 
\begin{equation}
    \Gamma_{\mathrm{ID2}}=n_N^{eq}\langle\sigma_{NN\rightarrow \phi_H}v_{\mathrm{rel}}\rangle,\quad \Gamma_{\mathrm{ID3}}=n_N^{eq,2}\langle\sigma_{NN\phi_L^{\dg}\rightarrow \phi_H}v_{\mathrm{rel}}^2\rangle \;,
\end{equation}

and the subscript number stands for the number of final bodies involved in a given process. Therefore, for $T<M_H$ and being $M_L<M_H$, \eqref{eq:inv_dec} implies that the inverse decays are less efficient than direct decays, so that wash-out processes can be neglected.

In this regime we can give a very simple estimate of the asymmetry that can be produced:

\begin{equation}
\label{eq:eta_asym}
    \eta_{\mathrm{DM}}\approx 2\frac{\epsilon}{g_*} \;,
\end{equation}

where $g_* \approx 10^2$ is the number of relativistic degrees of freedom, and the factor 2 comes from the fact that the three-body decay of $\phi_H$ has $\Delta D=6$ while the DM DCb carries $D=3$. From $\Omega_{\mathrm{DM}}\approx 5\Omega_b$ we then have:

\begin{equation}
    \Omega_{\mathrm{DM}}\propto \mtcb \eta_{\mathrm{DM}} \approx \frac{2\epsilon \mtcb}{g_*} \approx 5\eta_b m_p \;,
\end{equation}

so that enforcing $\mtcb\lesssim 75$ TeV translates in the following lower bound on $\epsilon$:

\begin{equation}
    \epsilon \gtrsim 3 \times 10^{-13} \;,
\end{equation}

which can be easily accomplished with perturbative couplings.\\
For scalars lighter than $\sim 10^{15}$ GeV, the asymmetry \eqref{eq:eta_asym} is generated after the freeze-out of the scalars and receives a Boltzmann suppression approximately given by $x_{\text{f.o.}}^{3/2}\exp(-x_{\text{f.o.}})$, where $x_{\text{f.o.}} = M_\phi/T_{\text{f.o.}}$ and $T_{\text{f.o.}}$ is the freeze-out temperature. Besides, the weak wash-out condition \eqref{eq:weakwashout} now reads:

\begin{equation}
   \left. \left(\frac{\Gamma_D}{2H}\right)\right|_{T=T_{\text{f.o.}}}<1
\end{equation}

This condition is equivalent to asking for the decay of the scalars to happen after their thermal freeze-out. As we shall see in the next Section, the largest asymmetry that we can get with heavy scalars is $\epsilon\approx 10^{-6}$. This implies that in order to satisfy the previous lower bound on $\epsilon$ we need $x_{\text{f.o.}}\lesssim 18$. For the scalars annihilating into dark gluons or DCquarks, together with the condition on the naturalness of the couplings and the weak washout condition, this requires $M_\phi\gtrsim 10^{10}$ GeV. 

\subsubsection{Computation of the asymmetry}\label{sec:asym_comp}
We are now ready to compute the asymmetry coefficient $\epsilon$, defined as:
\begin{equation}
\label{eq:epsilon}
    \epsilon = \frac{\Gamma(\phi_H\rightarrow \phi_L\dg NN)-\Gamma(\phi^\dagger_H\rightarrow \phi_L \overline{N}\overline{N})}{\Gamma_H}\;,
\end{equation}

where $\Gamma_H$ is the total decay rate of $\phi_H$. For simplicity, we shall take a very heavy scalar, such that \eqref{eq:eta_asym} holds. As discussed in Section \ref{sec:2phi}, the Lagrangian \eqref{eq:int_lag} allows for three different physical phases, all of which are expected to appear in the expression of the asymmetry. Indeed, in Appendix \ref{app_asym} we write the complete result for $\epsilon$ to show how the different phases contribute to the asymmetry. All the 1-loop computations have been carried out using \texttt{Package-X}\cite{Patel:2015tea,Patel:2016fam}, taking the massless fermion limit $m_N=0$. For simplicity, here we show the result obtained by fixing $\lambda_{LLL}=\lambda_{HLL}=0$, thus isolating a single CP-violating phase (see Appendix \ref{app_asym}). In particular, such choice of coupling selects only the contributions coming from the diagrams \ref{fig:loop_2} and \ref{fig:loop_5}.
The asymmetry \eqref{eq:epsilon} can be written as:

\begin{equation}
\label{asym_bm}
    \epsilon=-\frac{4}{\Gamma_H}\frac{1}{6}\frac{1}{2M_H}\frac{1}{2}\int\dPi_N\dPi_N\dPi_{\phi_L}\sum_{i,j}\Im[\mathcal{C}_{T_i}^*\mathcal{C}_{L_j}]\Im[\mathcal{A}_{T_i}^*\mathcal{A}_{L_j}]\;,
\end{equation}

where $\mathcal{A}_T$ and $\mathcal{A}_L$ are respectively the tree- and one-loop-level decay amplitudes, $\mathcal{C}_T$ and $\mathcal{C}_L$ the corresponding set of couplings, while $\Im$ denotes the imaginary part. Finally, the factor $\frac{1}{6}$ is the average over the initial dark color states. The sum runs over all possible diagrams.
Here

\[
\mathrm{d}\Pi\equiv g \frac{1}{(2\pi)^3}\frac{\mathrm{d}^3p}{2E}
\]

is the phase space measure. 
Under our simplifying assumptions we have:

\begin{equation}\label{eq:dasgupta_factor}
\begin{split}
\sum_{i,j}\Im[\mathcal{C}_{T_i}^*\mathcal{C}_{L_j}]&\Im[\mathcal{A}_{T_i}^*\mathcal{A}_{L_j}]=-\Im[\lambda_{LHH}^*\lambda_{HHH}y_L^*y_H]|y_H^2|\times\\
&\left(\frac{1}{(x-1)^2}\frac{M_L^2}{M_H^2-M_L^2}\frac{C_{\mathrm{DC}}}{16\pi}-\frac{C_{\mathrm{DC}}'}{x-1}\mathcal{D}_2^{HH}\left(\frac{p_{N1}}{M_H},\frac{p_{N2}}{M_H},\frac{M_L}{M_H}\right)\right)\;,
\end{split}
\end{equation}


where $x\equiv\frac{(p^\mu_{N1}+p^\mu_{N2})^2}{M_H^2}$, $C_{\mathrm{DC}}$ and $C_{\mathrm{DC}}'$ are dark color factors. The loop integral $\mathcal{D}_2^{HH}$ arises from the interference between the tree-level diagrams and the loop diagrams of Figure \ref{fig:loop_5} and is defined in Appendix \ref{app_asym}. The interference with diagrams \ref{fig:loop_2} is trivial and can be carried out analytically. Once that the integral in \eqref{asym_bm} is performed, we can compute the asymmetry generated in our model as a function of the different parameters.

\begin{figure}[H]
    \centering
    \includegraphics[width=\textwidth]{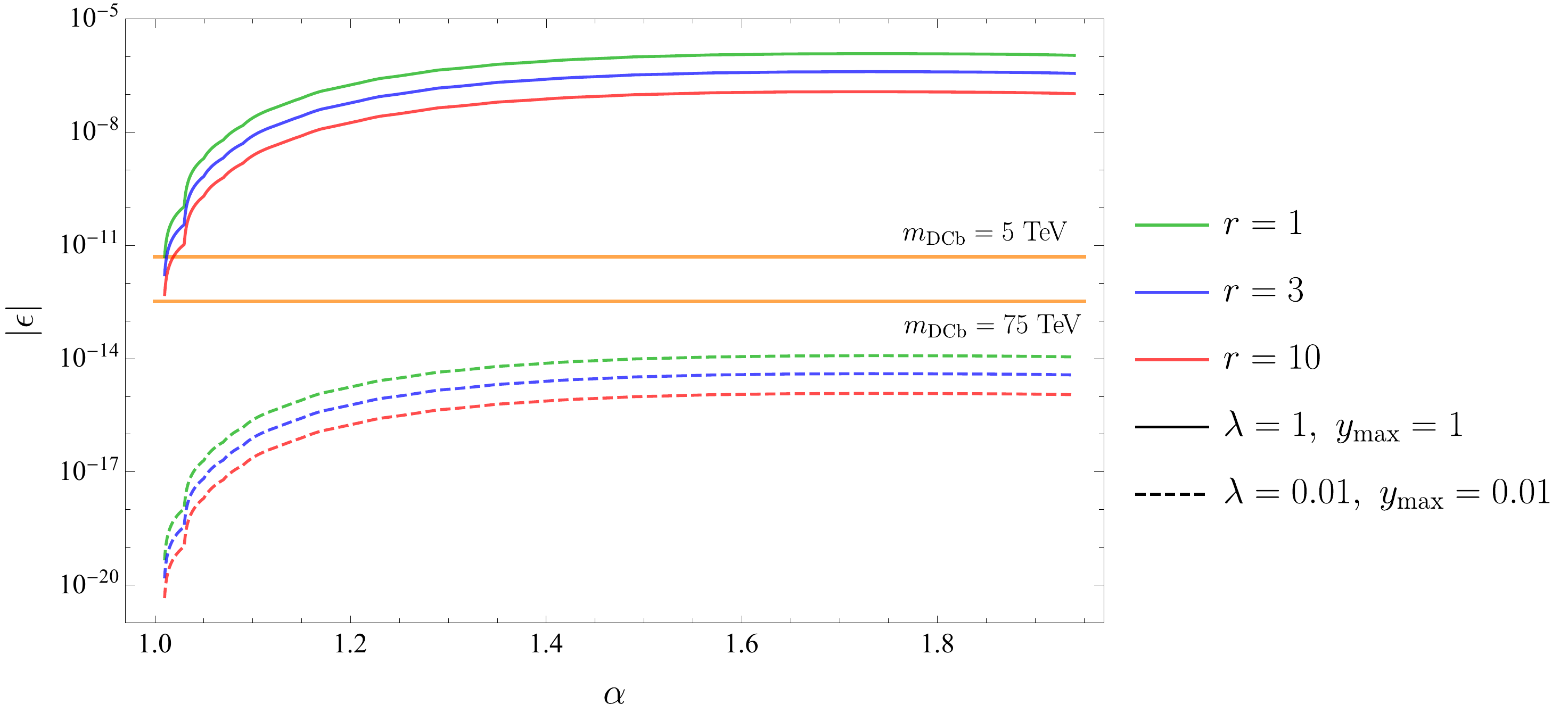}
    \caption{Asymmetry corresponding to several benchmark values of the parameters as a function of $\alpha = M_H/M_L$. Here we have taken $\lambda_{HHL}=\lambda_{HHH}=\lambda$ and $\text{arg}[\lambda_{LHH}^*\lambda_{HHH}y_L^*y_H]=\frac{\pi}{2}$ for simplicity. We have defined $r\equiv\left|\frac{y_L}{y_H}\right|$ while always keeping $y_{\max}\equiv\max(y_H,y_L)\leq 1$. Finally, the orange solid lines show the limiting values for $\epsilon$ coming from the equivalent limiting values on $\mtcb$.}
    \label{fig:asymmetries}
\end{figure}

In Figure \ref{fig:asymmetries}  we show the values of $\epsilon$ for some benchmark values of the relevant parameters. 

As we can see, in most of the cases we produce too much asymmetry and we need $\mathcal{O}(10^{-1})$ couplings to avoid the overclosure of the Universe. However, the plot is obtained by taking the largest possible CP-violating phase, so that we can gain some more parameter space by reducing its value. Moreover, this computation has been made for heavy scalars ($M_\phi\simeq 10^{15}$ GeV), with negligible wash-out processes and no Boltzmann suppression factor. If we take the scalar as light as $M_\phi\simeq 10^{10}$ GeV, instead, we get a suppression factor of order $10^{-6}$, as anticipated, and still produce the correct abundance for TeV scale masses of the DCb. Even lighter scalars would lead to an underproduction of the needed asymmetry for $\mtcb<75$ TeV. Finally, a more general analysis is shown in the scatter plot of Figure \ref{fig:asymmetries_scatter}, where all the phases have been included, as discussed in Appendix \ref{app_asym}, sampling them randomly. Similarly to Figure \ref{fig:asymmetries}, we have assumed all the couplings in the scalar potential to have a common absolute value $\lambda$. As we can see, the qualitative results do not change with respect to the simplified choice of couplings.

\begin{figure}[htp]
    \centering
    \includegraphics[width=0.9\textwidth]{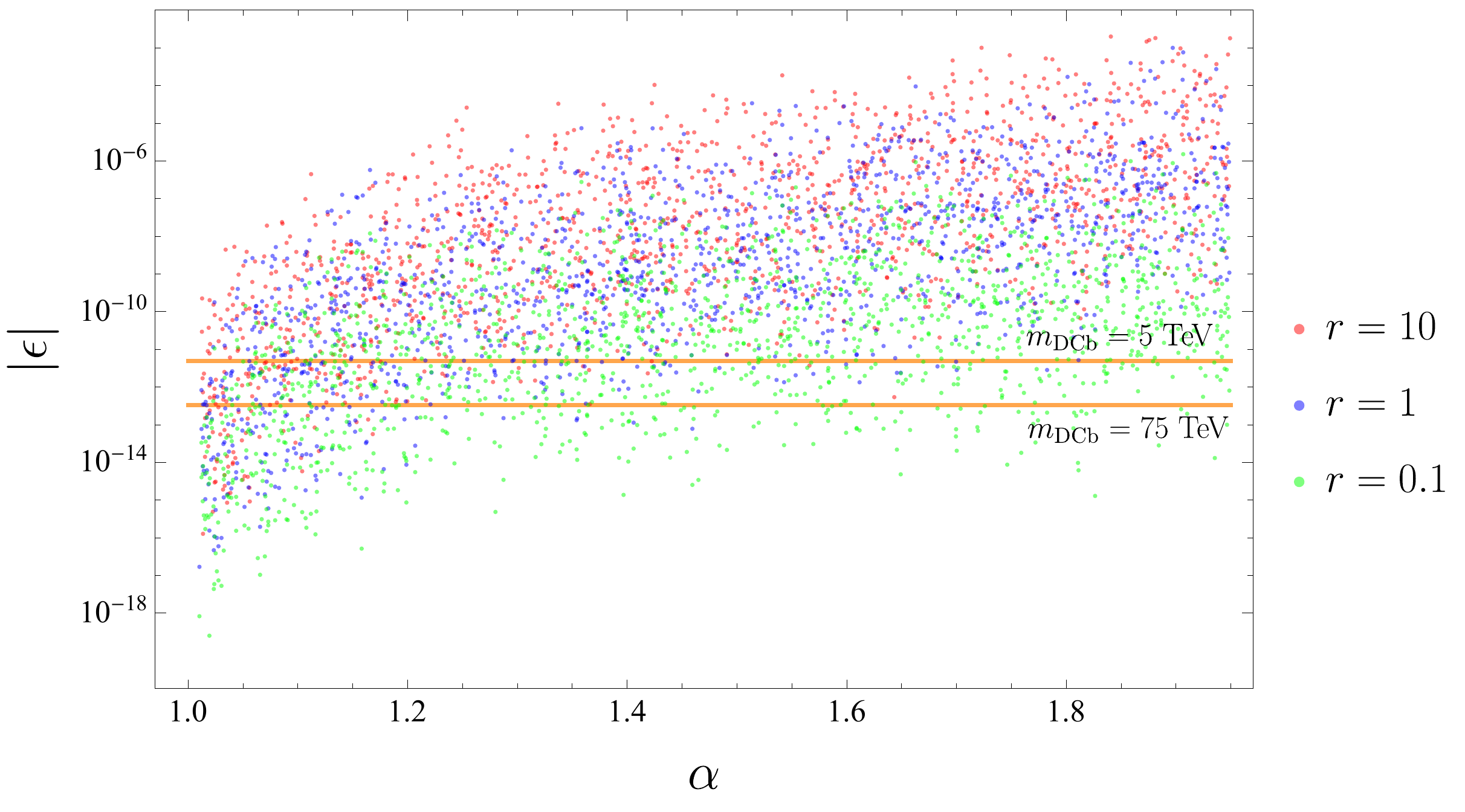}
    \caption{Asymmetry obtained by randomly sampling the values of $10^{-2}\leq \lambda, y_{\max} \leq 1$ and the three physical phases in $\left[-\pi,+\pi\right]$ (see Appendix \ref{app_asym}). The colors of the dots correspond to three different values of $r=\left|\frac{y_L}{y_H}\right|$.}
    \label{fig:asymmetries_scatter}
\end{figure}

\subsection{A single flavor of $\phi$}
In this attempt to generate the asymmetry, we introduce a single flavor of $\phi$.
This can be done either by exploiting the two decay channels of $\phi$.
\begin{equation}
    \phi \rightarrow \bar{N} \bar{N} \;, \quad \phi \rightarrow NNNN \;,
\end{equation}
or by exploiting some number-changing scattering processes.
Both approaches in the single flavor case suffer from the lack of couplings to form a phase-invariant complex combination. The only possibility is to allow different Yukawas $y_L,y_R$ for the left-handed and right-handed component of the DCquark $N$. However, this implies the presence of a chirality flip factor, which brings a suppression of order $m_N/M_\phi$. This is very small since $M_\phi \gtrsim 10^{15}$ GeV to ensure the weak washout condition for $\mathcal{O}(0.1)$ couplings, while is typically $m_\Psi \lesssim$ TeV for dark sectors enjoying an approximate chiral symmetry.
For simplicity we will only show the possibility of generating the asymmetry via scatterings, which was not considered in the previous section.
The processes contributing to leading order to the asymmetries are:

\begin{itemize}
    \item $\Delta\phi=2,$ $\Delta N=2$, $2\leftrightarrow 2$ processes
    \begin{equation}
    \label{2to2}
        \phi\phi\leftrightarrow NN,\quad \phi \overline{N}\leftrightarrow\phi\dg N \;.
    \end{equation}
    \item $\Delta\phi=1,$ $\Delta N=2$, $1\leftrightarrow 2$ processes
    \begin{equation}
    \phi\leftrightarrow \bar{N} \bar{N}\;.
    \end{equation}
\end{itemize}

Higher order processes are phase space and coupling suppressed. We expect the asymmetry to be first generated \textit{simultaneously} for $\phi$ and $N$ through the $2\leftrightarrow 2$ processes and then the asymmetry stored in $\phi$ is converted into an additional asymmetry in $N$ by its decay. In fact, if the $2\leftrightarrow 2$ produce, for example, a net abundance of $\phi\dg$ and $N$ over their antiparticles, that is if $\phi\phi\leftrightarrow NN$ and its cross-symmetric process dominate, then a larger number of $\phi\dg$ than $\phi$ will decay, thus increasing the asymmetry of the fermionic content of the dark sector. In order to write the Boltzmann equations, we shall write the amplitude for the first $2\leftrightarrow 2$ process in \eqref{2to2} as:

\begin{equation}
\begin{split}
    &|\mathcal{M}(\phi\phi\rightarrow NN)|^2=|\mathcal{M}(\overline{N}\overline{N}\rightarrow \phi\dg\phi\dg)|^2=(1+\epsilon)|\mathcal{M}_1|^2\\
    &|\mathcal{M}(\phi\dg\phi\dg\rightarrow \overline{N}\overline{N})|^2=|\mathcal{M}(NN\rightarrow \phi\phi)|^2=(1-\epsilon)|\mathcal{M}_1|^2
\end{split}
\end{equation}

The decay asymmetry is not independent of $\epsilon$ since CPT symmetry and unitarity require \cite{Kolb:1979qa}:

\begin{equation}
\begin{split}
    &\int\mathrm{d}\Pi_\phi\mathrm{d}\Pi_\phi|\mathcal{M}(NN\rightarrow \phi\phi)|^2 + \int\mathrm{d}\Pi_\phi|\mathcal{M}(NN\rightarrow \phi\dg)|^2 =\\
    &=\int\mathrm{d}\Pi_\phi\mathrm{d}\Pi_\phi|\mathcal{M}(\overline{N}\overline{N}\rightarrow \phi\dg\phi\dg)|^2 + \int\mathrm{d}\Pi_\phi|\mathcal{M}(\overline{N}\overline{N}\rightarrow \phi)|^2
\end{split}
\end{equation}
$\implies$
\begin{equation}
    \int\mathrm{d}\Pi_\phi|\mathcal{M}(NN\rightarrow \phi\dg)|^2 - \int\mathrm{d}\Pi_\phi|\mathcal{M}(\overline{N}\overline{N}\rightarrow \phi)|^2 = -2\epsilon\int\mathrm{d}\Pi_\phi\mathrm{d}\Pi_\phi|\mathcal{M}_1|^2 \;.
\end{equation}

For the same reason, the asymmetry related to the second $2\rightarrow 2$ process is related to that of a $2\rightarrow 3$ one, which we neglect.\\

\subsubsection{A quick estimate}
\label{quick_est}

We want to give an estimate of the amount of asymmetry that we can produce in our models with a single flavor resorting to the annihilation mechanism. The parameter $\epsilon$ can be estimated from the diagrams in Figure \ref{fig:22anni}, where the arrow indicates the particle number flow:

\begin{equation}\label{fig:22anni}
\vcenter{\hbox{\begin{tikzpicture}
  \begin{feynman}
    \vertex (i1){$\phi$};
        \vertex [below right = 1.7cm of i] (v1);
        \vertex [below left = 1.5cm of v1] (i2){$\phi$};
        \vertex [right = 1.5cm of v1] (v2);
        \vertex [above right = 1.5cm of v2] (f1){$N$}; 
        \vertex [below right = 1.5cm of v2] (f2){$N$};
        \diagram*[small]{(i) -- [charged scalar](v1) -- [anti charged scalar, edge label' = $\phi$](v2) -- [fermion](f1),
        (i2) -- [charged scalar](v1),
        (v2) -- [fermion](f2)};
  \end{feynman}
\end{tikzpicture}}}
\quad + \quad
\vcenter{\hbox{\begin{tikzpicture}
    \begin{feynman}
    \vertex (i1){$\phi$};
        \vertex [below right = 1.7cm of i] (v1);
        \vertex [below left = 1.5cm of v1] (i2){$\phi$};
        \vertex [right = 0.7cm of v1] (v3);
        \vertex [above right = 1cm of v3] (l1);
        \vertex [above = 0.08cm of l1] (n1) {$m_N$};
        \vertex [below right = 1cm of v3] (l2);
        \vertex [below = 0.1cm of l2] (n2) {$m_N$};
        \vertex [below right = 1cm of l1] (v4);
        \vertex [right = 0.7cm of v4] (v2);
        \vertex [above right = 1.5cm of v2] (f1){$N$}; 
        \vertex [below right = 1.5cm of v2] (f2){$N$};
        \diagram*[small]{(i) -- [charged scalar](v1) -- [anti charged scalar](v3) -- [insertion=0.99, fermion, quarter left](l1)  -- [fermion, quarter left](v4) -- [anti charged scalar](v2) -- [fermion](f1),
        (v3) -- [insertion=0.99,fermion, quarter right](l2) -- [fermion, quarter right](v4),
        (i2) -- [charged scalar](v1),
        (v2) -- [fermion](f2)};
    \end{feynman}
  \end{tikzpicture}}}
\end{equation}


The estimate roughly gives:
\begin{equation}
\label{eps_est}
    \epsilon\approx \frac{y^2}{16\pi^2}\frac{m_N^2}{M_\phi^2}\;,
\end{equation}

where $y=y_{L,R}$. The crucial point is that in order to get a net CP-violating phase in the diagrams, a chirality flip is needed to get both $y_L, y_R$ (notice that this is independent on whether the flavors circulating in the loop are one or two). This brings an extra suppression of order $m_N/M_\phi$ in the amplitude\footnote{This problem is also found when computing the asymmetry in the decay of a single scalar, leading to a further suppression of $\epsilon$.}. The DM asymmetry can be related to the SM one by the following estimate:

\begin{equation}
\label{eta_def}
    \eta_{\mathrm{DM}} \equiv \frac{\Delta n_{\mathrm{DM}}}{s}=\frac{5 m_p \eta_b}{\mtcb} \geq 6\times 10^{-15}\;,
\end{equation}

where $\eta_b = 10^{-10}$ is the baryon asymmetry and $\mtcb\lesssim 75$ TeV from the request of the symmetric component of the DM abundance being depleted by non-perturbative annihilations between baryons. Since the relevant dynamics for the asymmetry generation occurs in the nearby of $\phi$ freeze-out, $\eta_{\mathrm{DM}}$ can be estimated (neglecting wash-outs) as:

\begin{equation}
\label{eta_est}
    \eta_{\mathrm{DM}}\approx\epsilon Y_\phi^{f.o.} \approx \epsilon \frac{3.79 x_{fo}}{\sqrt{g_*} M_{\mathrm{P}} M_\phi \sigmaone}\;,
\end{equation}

where $x_{fo} = \frac{M_\phi}{T_{fo}}\approx 25$ and $\sigmaone\approx \frac{\lambda^2y^2}{M_\phi^2}$, where $\lambda$ is the dimensionless trilinear scalar coupling. Therefore the lower bound on $\eta_{\mathrm{DM}}$ can be converted into a bound into $M_\phi$ once we replace \eqref{eta_est} and \eqref{eps_est} into \eqref{eta_def}. Hence:

\begin{equation}
    M_\phi<9 \times 10^{12}~ \frac{1}{\lambda^2} \frac{m_N^2}{M_{\mathrm{P}}}=1.8\times 10^{-2}~ \frac{m_N}{\lambda^2} \lesssim \left(\frac{0.1}{\lambda}\right)^2~30\text{~TeV}\;,
\end{equation}

where in the last step we used $m_N<\ldc\lesssim \mtcb/\ndc \sim 25$ TeV. As we can see, we need rather small couplings to achieve freeze-out before confinement at temperatures near $\ldc$. Even with a very small $\lambda$, this minimal scenario suffers from strong wash-out processes. This can be seen in the Boltzmann equations, where some wash-out processes have the same rate as those responsible for the asymmetry generation. Therefore we expect the asymmetry in equation \ref{eta_est} to be completely wiped out, making the model not viable.\\

\section{Building the class 1 models: stable dark matter}\label{sec:buildingmodels}
In this section we will comment on a specific implementation of the mechanism described in Section \ref{sec:benchmarkmodel} in GC models in which DM is stabilized by a remnant discrete symmetry.
\subsection{Non-colored GC models}
We will analyze the original ACDM models whose DCquarks are not charged under $\mathrm{SU}(3)_c$. We will build models based on the $\phi^3$ term used in Section \ref{sec:benchmarkmodel}, and on a slight modification based on $\phi^4$ term.

\subsubsection{$\phi^3$}\label{sec:phi3}
Since the goal is to implement the benchmark model, we fix $\ndc=3$.
Notice that all the GC models whose DCquarks are not charged under $\mathrm{SU}(3)_c$ contain either the SM singlet $N$ or the $\mathrm{SU}(2)_L$ triplet $V$. Indeed if $\phi$ is a SM singlet, $V$ can be seen as three copies of $N$, leading to a factor of three in the total asymmetry. If the scalar $\phi$ couples to either one of them, the computation of the asymmetry presented in \ref{sec:asym_comp} follows immediately. If there are additional similar couplings to other DCquarks, we get an additional DCb number asymmetry. This happens whenever in the model are present a DCquark and its "tilded" SM-conjugate partner (like $L$ and $\tilde{L}$). This extra contribution to the asymmetry is numerically the same to the one computed in Section \ref{sec:asym_comp} (with the $\bar{N}\bar{N}$ finale states), differing only by some representation-dependent multiplicity coefficients. Each of these additional contributions are therefore expected to be equal to the asymmetry computed in the benchmark model, up to a $\mathcal{O}(1)$ coefficient.
Since the quantum numbers of $\phi$ are such that the only additional Yukawa is of the form $\phi \Psi \Psi$, the models enjoy a $\mathbb{Z}_2$ symmetry, under which only the DCquarks are non-trivially charged.
We summarize in table \ref{tab:complete_phi_3models} a possibile field content and the admitted extra Yukawa couplings for each of the original uncolored GC models. The $\phi^3$ term is always present to ensure the implementation of the mechanism.
\begin{table}[!htb]
    \centering
    \begin{tabular}{|c|c|c|}
        \hline
        Model & $\phi$ & Couplings\\
        \hline
        $V$ &  & $\phi VV$\\
        $N\oplus L$ &  & $\phi NN$ \\
        $N \oplus L \oplus \tilde{E}$ &  & $\phi NN$\\
        $V \oplus L $ &  & $\phi VV$\\
        $N \oplus L \oplus \tilde{L}$ & & $\phi NN, \phi L \tilde{L}$\\
        $V\oplus L \oplus N$ & $(\bar{6},1,1)_0$  & $\phi VV, \phi NN$\\
        $V \oplus L \oplus \tilde{E}$ &  & $\phi VV$\\
        $N \oplus L \oplus \tilde{L} \oplus \tilde{E}$ &  & $\phi NN, \phi L \tilde{L}$\\
        $L \oplus \tilde{L} \oplus E \oplus \tilde{E} \oplus N$ &  & $\phi NN, \phi E \tilde{E}, \phi L \tilde{L}$\\
        $N \oplus L \oplus \tilde{E} \oplus V$ &  & $\phi NN, \phi VV$\\
        \hline
         $V \oplus N$ & $(\bar{6},1,5)_0$ & $\phi VV$\\
        \hline
    \end{tabular}
    \caption{List of non-colored GC models, $\phi$ representation under $\left( \sut, \mathrm{SU}(3)_c,\mathrm{SU}(2)_L \right)_Y$ and allowed Yukawa couplings (in addition to the ones of the original models with $H$). The models are based on a $\sut$ gauge group and $\phi^3$ $\udb$-breaking term.}
    \label{tab:complete_phi_3models}
\end{table}\\
The only model in which a different SM representation for $\phi$ is needed to forbid the existence of multiple couplings is $V \oplus N$. This is the only GC model in which there are 2 extra unbroken $\mathrm{U}(1)$, the $V$ and $N$ species numbers. Each coupling with $\phi$ breaks one of them, or a combination of the two.
Since these symmetries are only broken by the heavy $\phi$, below $M_\phi$ the effective theory will enjoy the extra $\mathrm{U}(1)$ symmetries: afterall, the effective theory coincides with the original GC model in this regime. This implies that in general the model can possess a shared asymmetry between the two species in the IR. By properly choosing the SM representation of $\phi$, the scalars can be taken to couple to a single species, which for concreteness we will take to be $V$:
\begin{equation*}
\mathcal{L} \supseteq \phi V V\
\end{equation*}
In this case, even in the UV theory the $N$ species number is unbroken, and therefore the dark sector will contain an asymmetric, non-thermally generated component stored in $V$ DCquarks, and a symmetric, thermally produced component stored in the $N$ DCquarks. Below confinement, the \tcp decay via the non-renormalizable 5d operators, leaving only the DCbs. By taking $\mtcb\lesssim 75$ TeV, as shown in Section \ref{sec:annihilation}, only the asymmetric part survives. It is stored in the form of DCbs containing $V$ as valence DCquark. Notice however that given the symmetries of the IR theory, the asymmetry will in general be shared among the different type of DCbs containing $V$.
If instead the $\phi$ couples to both species, like:
\begin{equation}\label{eq:speciesbreakingphi}
    \mathcal{L}\supseteq y_N\phi NN + y_V \phi VV \;,
\end{equation}
both species number will be broken.
Virtual $\phi$ exchange will mediate the conversion of DCbs with different species number, together with a DC$\pi$ emission (if kinematically allowed) or SM gauge boson radiation. For the sake of showing what can go wrong if the coupling in equation \ref{eq:speciesbreakingphi} is not forbidden, we assume a splitting $m_V-m_N\simeq \ldc$. The rate of species conversion followed by a \tcp emission can be estimated roughly as
\begin{equation}
\Gamma \approx \frac{y^2}{8 \pi}\left(\frac{\mtcb}{M_\phi}\right)^4\ldc \approx \frac{1}{\ndc}\left(\frac{y}{0.1}\right)^2\left(\frac{\mtcb}{30 \;\mathrm{TeV}}\right)^5\left(\frac{10^{15}\;\mathrm{GeV}}{M_\phi}\right)^4 10^{-46} \;\mathrm{GeV} \;.
\end{equation}
Since DC$\pi$ quickly decay after production via the $M_{\mathrm{P}}$-suppressed 5D operator, they decay shortly after the species conversion of the DCbs, as shown in Section \ref{sec:higherdim}. The same is true for direct gauge boson emission. 
For generic values of the coupling we expect $N$ and $V$ asymmetries to be of the same order, and therefore the DM abundance is roughly equally stored in the heavier and lighter DCbs. For this reason the fraction of DM energy injected in the SM is expected to be roughly proportional to the splitting-DCb mass ratio, which in this example is $\mathcal{O}(1)$.
Such late energy injection is in tension with bounds coming from ID experiments \cite{Cohen:2016uyg},\cite{Slatyer:2017sev}:
\begin{equation}
    \Gamma < 10^{-53} \; \mathrm{GeV} \;.
\end{equation}
Such constrain can be evaded by ad hoc model building, \emph{e.g.} by tuning the splittings or by avoiding the simultaneous coupling of $\phi$ to both DCquarks, as accomplished with the representation presented in table \ref{tab:complete_phi_3models}.
In the other models, even if the DCb species conversion is kinematically allowed, it happens shortly after $\ldc$ via renormalizable Yukawas with $H$, rather than with the different $\phi$ couplings. Therefore there are no dangerous late time \tcp decays and no further restrictions.
\subsubsection{$\phi^4$}\label{sec:phi4}
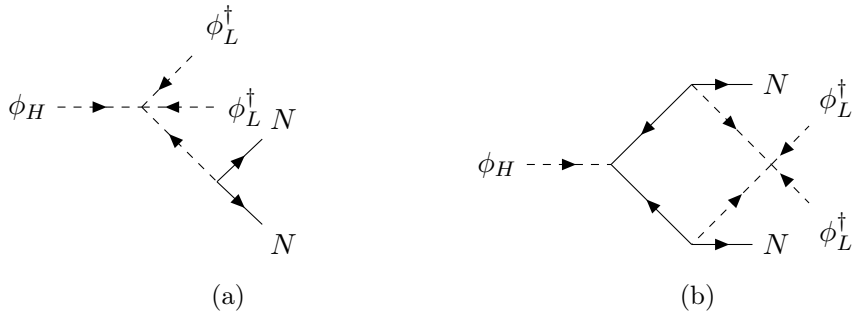
\begin{figure}[H]
    \centering
    \begin{subfigure}[b]{0.4\textwidth}
    \begin{tikzpicture}[/tikzfeynman/small]
        \begin{feynman}
        \vertex (i){$\phi_H$};
        \vertex [right = 1.5cm of i] (v1);
        \vertex [above right = 1cm of v1] (s1){$\phi^\dagger_L$};
        \vertex [right = 1cm of v1] (s2){$\phi^\dagger_L$};
        \vertex [below right = 1.4cm of v1] (v2);
        \vertex [above right =0.8cm of v2] (f1){$N$};
        \vertex[below right =0.8cm of v2] (f2){$N$};
        \diagram*[small]{(i) -- [charged scalar](v1) -- [anti charged scalar](s1),
        (v1) -- [anti charged scalar](s2),
        (v1) -- [anti charged scalar](v2)--[fermion](f1),
        (v2)--[fermion](f2)};
        \end{feynman}
    \end{tikzpicture}
    \caption{}
    \end{subfigure}
    \begin{subfigure}[b]{0.4\textwidth}
    \begin{tikzpicture}[/tikzfeynman/small]
        \begin{feynman}
        \vertex (i){$\phi_H$};
        \vertex [right = 1.5cm of i] (v1);
        \vertex [above right = 1.5cm of v1] (w1);
        \vertex [below right = 1.5cm of v1] (w2);
        \vertex [right = 2.11cm of v1] (w3);
        \vertex [right = 0.8cm of w1] (f1){$N$};
        \vertex [right = 0.8cm of w2] (f2){$N$};
        \vertex [above right = 0.7cm of w3] (f3){$\phi\dg_L$};
        \vertex [below right = 0.7cm of w3] (f4){$\phi\dg_L$};
        \diagram*[small]{(i) -- [charged scalar](v1) -- [anti fermion](w1) -- [charged scalar](w3) -- [anti charged scalar](w2) -- [fermion](v1),
        (w3) -- [anti charged scalar](f3),
        (w1) -- [fermion](f1),
        (w2) -- [fermion](f2),
        (w3) -- [anti charged scalar](f4)};
        \end{feynman}
    \end{tikzpicture}
    \caption{}
    \end{subfigure}
\caption{Tree level process (left) and an example of a 1-loop correction (right) in the $\phi^4$ models.}
    \label{fig:phi4process}
\end{figure}

In principle, it's possible to pick $\mathrm{SU}(4)$ as dark color gauge group, and to have a quartic potential term
\begin{equation}
V=\lambda \epsilon^{ijkl}\epsilon^{i'j'k'l'}\phi_{ii'}\phi_{jj'}\phi_{kk'}\phi_{ll'} \;,
\end{equation} 
with $\phi$ in the 10-dimensional symmetric representation of $\mathrm{SU}(4)$. This term substitutes the cubic term of the benchmark model, while the Yukawas with the DCquarks are the same. Notice that here the coefficient of the quartic term is automatically dimensionless, and it's $\mathcal{O}(1)$ in a natural theory.
The difference with the mechanism of the benchmark model is that now the secondary decay channel for the heaviest scalar is a 4-body decay: $\phi_H \rightarrow \phi^\dagger_L \phi^\dagger_L N N $.
In Figure \ref{fig:phi4process} we show the tree level expression and an example of loop level contribution.

For simplicity, we will not do the explicit computation, but we expect the generated asymmetry to be suppressed by an additional phase-space factor, leading even in this case to the correct amount. \\
It's interesting to see that even in this $\mathrm{SU}(4)$ scenario, there is still an accidental remnant $\mathbb{Z}_8$ symmetry, whose charges are listed in table \ref{tab:z8}.
\begin{table}[!htb]
    \centering
    \begin{tabular}{|c|c|}
    \hline
        Field & $\mathbb{Z}_8$ \\
        \hline
        $\phi$ & $\omega^2$ \\
        \hline
        $\Psi$ & $\omega$\\
        \hline
        SM & 1\\
        \hline
    \end{tabular}
    \caption{$\mathbb{Z}_8$ charges of the $\mathrm{SU}(4)$ model, where $\omega = e^{i\frac{\pi}{4}}$.}
    \label{tab:z8}
\end{table}
DCbs, being made by four generic DCquarks $\Psi$, carry $-1$ charge under this $\mathbb{Z}_8$, making the lightest DCb stable.
\subsection{Colored GC models}
In this scenario it is not possible to directly use the benchmark model. Indeed the consequence of the cubic term in the potential is to force the hypercharge of $\phi$ to be $0$. This hypercharge assignment excludes the possibility of a Yukawa between $\phi$ and the two colored DCfermions, as seen from the field content of colored ACDM models in Appendix \ref{sec:goldenclass}. The need to have the Yukawa $\phi \Psi \Psi$ then forces $\phi$ to have non-zero hypercharge, and therefore to resort to a different potential term.
\subsubsection{$\phi^3 H^*$}\label{sec:phi3h}
The only other way to write a potential term that violates the $\udb$ charge for $\phi$ (fixed by the Yukawa), and that can be used to mediate the decay of $\phi$, is to involve an additional light scalar. If we stick to a single representation for $\phi$, the only other light scalar is the Higgs field\footnote{and possibly its $\mathrm{SU}(5)$ partner, that we will not consider.}. This forces the gauge group to be $\sut$.
If $\phi \in (\bar{6},\bar{6},2)_{1/6}$ of $\sut \times G_{\mathrm{SM}}$\footnote{This SM representation cannot fit in any SU(5) multiplet. A possible choice compatible with the SU(5) embedding is $(\bar{6},3,2)_{1/6}$. In this case, we need at least 3 scalar flavors to write the U(1)$_{\mathrm{DB}}$-violating potential.}, we can build the following Yukawa:
\begin{equation}
   \phi_{ij,ab}Q^{ia}\tilde{D}^{jb},
\end{equation}
and the following $\udb$-violating potential term:
\begin{equation}
    \lambda \epsilon^{abc}\epsilon^{a'b'c'}\epsilon^{ijk}\epsilon^{i'j'k'}\phi_{ii'aa'}\phi_{jj'bb'}\phi_{kk'cc'}H^* \;,
\end{equation}
where $i,j,k$ and $a,b,c$ are dark color and color indices respectively (the $\mathrm{SU}(2)_L$ contraction is left implicit). 
No other terms are allowed by gauge invariance at the renormalizable level. Notice that the coefficient in the potential is dimensionless, like in $\phi^4$ models of Section \ref{sec:phi4}.
The DCbs are stabilized by the same $\mathbb{Z}_2$ of $\phi^3$ models, under which only the DCquarks are charged.
Like for the models described in Section \ref{sec:phi4} in this scenario the secondary decay of the heaviest scalar is a 4-body process: $\phi_H \rightarrow \phi^\dagger_L Q \tilde{D} H$. A difference is that here three out of four of the final-state particles are ultrarelativistic (the DCquarks and $H$), making it more similar to the process described in Section \ref{sec:benchmarkmodel}.
In Figure \ref{fig:phi3hprocess} we show the tree level secondary decay channel and one of its 1-loop corrections.
\begin{figure}[H]
    \centering
    \begin{subfigure}[b]{0.4\textwidth}
    \begin{tikzpicture}[/tikzfeynman/small]
        \begin{feynman}
        \vertex (i){$\phi_H$};
        \vertex [right = 1.5cm of i] (v1);
        \vertex [above right = 1cm of v1] (s1){$\phi^\dagger_L$};
        \vertex [right = 1cm of v1] (s2){$H$};
        \vertex [below right = 1.4cm of v1] (v2);
        \vertex [above right =0.8cm of v2] (f1){$Q$};
        \vertex[below right =0.8cm of v2] (f2){$\tilde{D}$};
        \diagram*[small]{(i) -- [charged scalar](v1) -- [anti charged scalar](s1),
        (v1) -- [scalar](s2),
        (v1) -- [anti charged scalar](v2)--[fermion](f1),
        (v2)--[fermion](f2)};
        \end{feynman}
    \end{tikzpicture}
    \caption{}
    \end{subfigure}
    \begin{subfigure}[b]{0.4\textwidth}
    \begin{tikzpicture}[/tikzfeynman/small]
        \begin{feynman}
        \vertex (i){$\phi_H$};
        \vertex [right = 1.5cm of i] (v1);
        \vertex [above right = 1.5cm of v1] (w1);
        \vertex [below right = 1.5cm of v1] (w2);
        \vertex [right = 2.11cm of v1] (w3);
        \vertex [right = 0.8cm of w1] (f1){$Q$};
        \vertex [right = 0.8cm of w2] (f2){$\tilde{D}$};
        \vertex [above right = 0.7cm of w3] (f3){$\phi\dg_L$};
        \vertex [below right = 0.7cm of w3] (f4){$H$};
        \diagram*[small]{(i) -- [charged scalar](v1) -- [anti fermion](w1) -- [charged scalar](w3) -- [anti charged scalar](w2) -- [fermion](v1),
        (w3) -- [anti charged scalar](f3),
        (w1) -- [fermion](f1),
        (w2) -- [fermion](f2),
        (w3) -- [scalar](f4)};
        \end{feynman}
    \end{tikzpicture}
    \caption{}
    \end{subfigure}
\caption{Tree level process (left) and an example of a 1-loop correction (right) in the $\phi^3H^*$ models.}
    \label{fig:phi3hprocess}
\end{figure}
We expect that the asymmetry computed in this scenario is simply suppressed by an additional $\sim1/(16\pi^2)$ phase space factor with respect to the one computed for the benchmark model in Section \ref{sec:asym_comp}.  By comparing with the results of Figure \ref{fig:asymmetries} we expect that even in the presence of the additional phase space the correct amount of asymmetry can still be produced.\\
If $\phi \in \left(3,3,2\right)_{1/6}$, we could have built the same lagrangian terms, however the presence of the mixed Yukawa $\phi^\dagger\bar{l} \tilde{D}$ would have made the DM unstable, although sufficiently long-lived to satisfy current bounds, as explored in Section \ref{sec:othermodels}.
\subsection{Higher dimensional operators}\label{sec:higherdim}
So far, we have studied the stability of the DM candidate by considering only renormalizable operators made by SM fields, DCquarks and $\phi$. However, this is not enough to study the DM stability.
Indeed the Lagrangian at the cutoff of the theory $\lcut$ (not to be confused with $\luv=M_\phi$, the cutoff of the original ACDM models) might contain already dangerous non-renormalizable operators that can mediate fast decay.
If a $d$-dimensional operator mediates DCb decay, the lifetime of the DCb is estimated using naive dimensional analysis:
\begin{equation}\label{eq:nr_lifetime}
\Gamma \simeq \frac{1}{8\pi}\left(\frac{\mtcb}{\lcut}\right)^{2(d-4)}\mtcb \le 10^{-53}\; \mathrm{GeV} \;,
\end{equation}
where the inequality comes from ID bounds on DM decay.\\
A possible natural cutoff for ACDM models is the GUT scale: at this scale the GUT completion is expected to complete SM fields and DCquarks to $\mathrm{SU}(5)$ representations. The particular value at which $\mathrm{SU}(5)$ GUT is achieved is model dependent \cite{Antipin:2015xia}: since by naturalness we want it to be slightly heavier than the dark scalars $\phi$, we will take $M_{\mathrm{GUT}}\gtrsim 10^{15}$ GeV, compatible with the typical values allowed by proton stability\footnote{Depending on the value of the coupling $\alpha_{\mathrm{GUT}}$, proton stability might require $M_{\mathrm{GUT}}\gtrsim 10^{16}$.}.
In order to check that the GUT completion does not affect the DCb decay, we assume that the only additional fields that are present at the GUT scale (but possibly appearing even at lower scales) are only the GUT partners of the Higgs and DCquarks, the $\mathrm{SU}(5)$ Higgs, and the heavy $\mathrm{SU}(5)$ bosons. It's important for the argument that the only non-trivial $\sun$ representations of extra fields, both light and heavy, are the (anti)fundamental for fermions and the 2-symmetric for the scalars.
In this case, even above GUT scale all the arguments given in Section \ref{sec:buildingmodels} hold, since the absence of the symmetry-breaking mixed Yukawa was based not on SM-charge arguments, rather on $\sun$ invariance alone. Under this assumption, integrating out fields at $M_{\mathrm{GUT}}$ does not generate any symmetry-breaking term. \\
Notice that the argument holds for other possible cutoff scales of the theory, provided the condition on $\sun$ charges of the new fields is satisfied. In particular, adding a family of total singlet right handed neutrinos does not spoil this picture. Therefore it's possible to implement the asymmetry generation in the SM via the thermal leptogenesis mechanism without ruining DM stability.\\
Adding different $\sun$ representations can potentially make the model of the classes discussed in Section \ref{sec:othermodels} unviable, with DM decaying too fast unless $\lcut \sim M_{\mathrm{P}}$. \\
The next cutoff of the theory is the Planck mass $M_{\mathrm{P}}$, at which a quantum theory of gravity completes the IR QFT.
In GC models in which \tcp do not decay via the Higgs Yukawa (like $V, V\oplus N$ models) 5d operators of the form
\begin{equation}
    H^\dagger H \left( \mathrm{DC}\pi \right) \;, \quad H^\dagger \sigma^a H \left( \mathrm{DC}\pi \right)^a
\end{equation} offer a fast decay channel into SM particles, as seen from equation \ref{eq:nr_lifetime}\footnote{There are also lagrangian terms made by integrating out the renormalizable operators, but they might be too much suppressed by the $M$ scale appearing in the virtual line. Planck-suppressed 5d operators offer a model independent way to make the DC$\pi$ decay fast.}. The effect of these operator on DCb decay must be checked.\\
Taking $\mtcb \sim 30 $ TeV, $\lcut\sim M_{\mathrm{P}}$, it follows from equation \ref{eq:nr_lifetime} that only 5d operators can potentially mediate DCb fast decays.
Therefore, if the residual discrete symmetry exists  at $d=5$, the DCb will be stable for all practical purposes. We remark that the existence up to 5d of the stabilizing symmetry is a sufficient, but not necessary condition: indeed even if a symmetry-breaking operator existed, the operator mediating DCb decay might have further suppression factors due to the integration of additional heavy fields like $\phi$. In the next sections we will show that the models presented keep their discrete symmetry even at 5d.
\subsubsection{$\phi^3$ models}
For these models $\sun=\sut$, $\phi$ is in the $\bar{6}$ of $\sut$, and the symmetry is a $\mathbb{Z}_2$ symmetry under which only the DCquarks are charged.
Operators that break the $\mathbb{Z}_2$ must contain on odd number of DCfermions.
The only kind of 5d operator containing an odd number of DCquarks is schematically $SSS \psi \Psi$ where $S=\phi,H$ is a generic scalar.
Invariance under $\sut$ forces one or two $\phi$.
This is forbidden again by $\sut$ for the particular choice of representation of $\phi$, therefore such operators do not exist.
\subsubsection{$\phi^3 H^*$}
We have presented a $\sun=\sut$ model with $\phi \in \left(\bar{6},2 \right)_{1/6}$.
Exactly as for $\phi^3$ models, the fact that $\phi$ is in the 2-symmetric representation of dark color prevents the construction of $\sut$ invariant operators with a single DCquark. The symmetry is therefore preserved at 5d level.\\
On a side note, the SM baryon number, under which only the SM quarks (and eventually the $\mathrm{SU}(5)$ Higgs partner) are charged, is not violated even at 5d, thanks to the specific $\sun$ and $\mathrm{SU}(3)_c$ representations of $\phi$. 
\subsubsection{$\phi^4$}
In this scenario the symmetry stabilizing the DCb is a $\mathbb{Z}_8$, under which both $\phi, \Psi$ are charged, as shown in table \ref{tab:z8}. This implies that there are additional symmetry-breaking operators that must be checked with respect to the previous cases.
However, $4$-ality considerations tell us that in order to preserve $\mathrm{SU}(4)_{\mathrm{DC}}$ invariance, the only possible dark field combinations appearing in such operator are either $\bar{\Psi} \Psi$, which does not break the stabilizing symmetry, $\phi \Psi \Psi$, which  does not break by construction the symmetry (it's the same combination appearing in the dark Yukawa), $\phi^4$, which again does not break the symmetry\footnote{$\phi^2$ does not appear since it's built by antisymmetrizing the symmetric indices of $\phi$. For the same reason, no $\phi^\dagger \Psi \Psi$ term is possible.}. Therefore also this class of models is protected from 5d operators.
\subsection{DCb-$\overline{\mathrm{DCb}}$ oscillations}\label{sec:tcboscillation}
\begin{figure}
    \centering
    \includegraphics[width=0.45\textwidth]{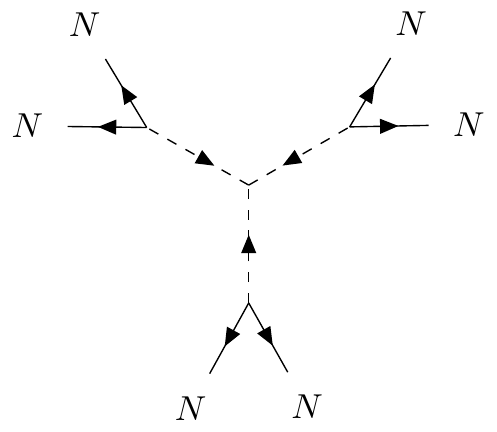}
    \caption{UV diagram leading to the DCb-$\overline{\mathrm{DCb}}$ oscillations in the IR phase.}
    \label{fig:oscillation}
\end{figure}

A feature of the models presented in this section is that the symmetry that stabilizes the DCb below the confinement scale is a $\mathbb{Z}_2$ under which the DCb is non-trivially charged. However, this implies that also its conjugate, $\overline{\mathrm{DCb}}$, carries the same charge. Therefore, oscillations between the two are not forbidden by any quantum number that was protecting the DCb from decaying. A symmetry that would prevent this to happen is a symmetry that admits complex charges for the DCb (like $\mathbb{Z}_3$ or $\mathbb{Z}_4$). However, this is not what happens in the models under scrutiny. In a certain sense, our models have a "minimally" asymmetric DM candidate, given that the only property that distinguishes between the DM and its conjugate is the complex representation of the dark color gauge group, which is confined below $\ldc$.
The presence of oscillations can significantly alter the cosmological history of the DM, regenerating the symmetric component, washing-out the asymmetry and possibly recoupling the annihilation process between the DCb and its conjugate \cite{PhysRevLett.108.011301},\cite{Cirelli:2011ac}. It's paramount to understand if they happen in the models we are considering.
In the IR EFT the oscillations are due to a "Majorana" mass term that depends on whether the DCb is a fermion or a scalar, which in turn depends on the number of its constituent DCquarks:
\begin{equation}\label{eq:IR_mass}
    \mathcal{L}\supseteq c \mtcb \mathcal{B} \mathcal{B} \quad(\ndc \;\mathrm{odd})  \;,\qquad \mathcal{L}\supseteq c \mtcb^2 \mathcal{B} \mathcal{B} \quad(\ndc \;\mathrm{even})\;,
\end{equation}
where $\mathcal{B}$ is a generic DCb, $c$ a dimensionless constant, and we assume that the scale at which the term is generated is around the confinement scale $\ldc$. For simplicity, we will work only the benchmark model of Section \ref{sec:benchmarkmodel}. In this case $\ndc=3$, and the DCb is a fermionic candidate. For the $\ndc=4$ case we expect a higher suppression of the oscillation-mediating operator due to its higher dimensionality. In order to show the role of the number of color, we will write the formualae without setting $\ndc$ to its value of $3$, and only setting it at the end to make the estimate of the mass splitting in our specific models. In a generic ACDM model the term in equation \ref{eq:IR_mass} can mix different DCb species, but this remark will not affect future conclusions about oscillations.\\
We have to estimate $c$ in equation \ref{eq:IR_mass}. The term in the UV theory that interpolates the Majorana mass term is:
\begin{equation}\label{eq:uvmajorana}
    \mathcal{L}\supseteq \frac{\lambda}{M_\phi^{3\ndc-4}}(N)^{\ndc}(N)^{\ndc}\;,
\end{equation}
where as we have seen, $\ndc=3$ for $\phi^3$ models\footnote{The $\phi^3H^*$ model leads to a "Majorana term plus Higgs".}. Since the operator is generated by integrating out the heavy scalars of mass of order $M_\phi$, this scale appears in the effective operator in equation \ref{eq:uvmajorana}. In particular, the diagrams that generate this term are in Figure \ref{fig:oscillation}.

This leads to an estimate for $c$:
\begin{equation}
c \sim k \lambda y^{N_\mathrm{DC}}\left(\frac{\mtcb}{M_\phi}\right)^{3\ndc-4}\;,
\end{equation}
where the $M_\phi$ power have been set by the previous argument, while the $\mtcb$ power comes from dimensional analysis and the fact that it is the relevant scale to use when interpolating the DCb with the DCquarks. Here $\lambda$ is the dimensionless coupling appearing in the cubic term of the potential $\lambda M_\phi \phi^3$ for the $\phi^3$ models.
The coefficient $k$ is non-perturbative in nature and we will assume that is $\mathcal{O}(1)$.
For the $\phi^3H^*$ model, the Majorana term is only obtained when the Higgs gets a vev, leading to a further suppression factor $v/M_\phi$ with respect to the pure $\phi^3$ case. 
This small mixing term leads to a splitting in mass between the two mass eigenstates of the DCb-$\overline{\mathrm{DCb}}$ system. 
The mass splitting comes from diagonalizing the matrix:
\begin{equation}\label{eq:mass_matrix}
    \begin{bmatrix}
    \mtcb & \mtcb \frac{\lambda y^{\ndc}}{2} \left(\frac{\mtcb}{M_\phi}\right)^{3\ndc-4}\\
    \mtcb \frac{\lambda y^{\ndc}}{2} \left(\frac{\mtcb}{M_\phi}\right)^{3\ndc-4} & \mtcb
    \end{bmatrix}
\end{equation}
The splitting is approximately proportional to the ratio of the off diagonal term and the diagonal one times the common mass $\mtcb$, in the limit of small mixing (satisfied since $M_\phi \gg m_{\mathrm{DCb}}$).
Therefore we have that the splitting is:
\begin{equation}
    \Delta m \approx \lambda y^{\ndc} \mtcb\left(\frac{\mtcb}{M_\phi} \right)^{3\ndc-4}\;.
\end{equation}
For the $\sut$ $\phi^3$ models, we have\footnote{For the $\phi^3 H^*$ models there is the additional $v/M_\phi$ suppression}.:
\begin{equation}\label{eq:mass_splitting}
    \Delta m \approx 7\lambda y^3 \left(\frac{\mtcb}{30 \mathrm{TeV}}\right)^6 \left( \frac{10^{16} \mathrm{GeV}}{M_\phi}\right)^5 10^{-49} \; \mathrm{GeV}\;.
\end{equation}
Sizable oscillations (of the order of the initial asymmetric abundance) can begin only if they are faster than the Hubble rate $\hub$ \cite{PhysRevLett.108.011301, Tulin:2012re}: $\Delta m \gtrsim \hub$. Since the present value of $\hub$ is $10^{-42}$ GeV, it's possible to avoid this constrain by taking $\mtcb$ in the multi-TeV range (as required by the constrain from elimination of the symmetric component and collider bounds), and $M_\phi$ around the GUT scale. In this case the DM candidate will not oscillate for all the practical purposes, and the asymmetry will be preserved. In the $\phi^3 H^*$ and $\phi^4$ models the oscillations are further suppressed, therefore this effect cannot be seen at the current stage of cosmological history even in the other models presented.\\
However, in some of our models, even in the presence of oscillations, washout can be avoided due to the peculiarity of the IR dynamics.
For concreteness, we will consider the $N_\mathrm{DC}=3$, $\Psi=V$ model, in which the DCquark is a $\mathrm{SU}(2)_L$ triplet. The DCb is itself a $\mathrm{SU}(2)_L$ triplet, and it can form dark nuclei \cite{Redi:2018muu}, as will be shown in Section \ref{sec:pheno}.
For TeV scale masses of the DCb, roughly up to $\mathcal{O}(1)$ fraction of the DCbs gets bound in a deuterium-like bound state, while much less for 100 TeV DCbs.
DCbs inside a nucleus experience a different potential with respect to free DCbs, shifting the diagonal upper and lower entries of the mass matrix in Eq. \ref{eq:mass_matrix} by $\mp E_n$ respectively.
The same is true also in the SM for neutrons and antineutrons, where the energy difference between the two particles inside a nucleus is $E_{\bar{n}}-E_n\simeq 100 \; \mathrm{MeV}$ \cite{Mohapatra:2009wp}.
The energy splitting is due to the strong nuclear potential, and so to the $\mathrm{SU}(3)$ structure of QCD. Therefore, a similar result apply to our case, and we will assume, for practical purpose, the energy difference inside a dark nucleus between the two DCbs to be of order $\ldc$.
By following the arguments in \cite{Cui:2012jh}, the oscillation rate inside a nucleus  not only is slowed  by the larger mass splitting induced by the nuclear potential, but it gets damped by scattering events between the constituents happening at a rate proportional to the inverse size of the nucleus $\ldc$. This leads to a suppression of the oscillation rate by an extra factor
\begin{equation}
    \frac{\Delta m}{\ldc}= N_\mathrm{DC} \left(\frac{m_{\mathrm{DCb}}}{M_\phi}\right)^5\simeq 0.7 \times 10^{-53}\left(\frac{m_{\mathrm{DCb}}}{30 \mathrm{TeV}}\right)\left( \frac{10^{15}\mathrm{GeV}}{M}\right) \;,
\end{equation}
forbidding oscillations even for lighter $\phi$. So, for $M_\phi\lesssim 10^{14}\; \mathrm{GeV}$ it's possible to regenerate a symmetric component of the unbounded DCbs, while preserving the asymmetric component intact and stored in the dark nuclei, avoiding the washout of this component.
Notice that this scenario is peculiar to our models: it needs the DCbs to be able to form dark nuclei and a small Majorana mass for the DCbs. The presence of a symmetric component leads to residual annihilations that can in principle be tested using Indirect Detection experiments. More details of this scenario are further discussed in Section \ref{sec:pheno_osc}.
\section{Other models}\label{sec:othermodels}
In Section \ref{sec:buildingmodels} we have shown models predicting stable DM.
For colored GC models, restricting to stable DM, this comes at the price of modifying the mechanism for asymmetry generation presented in \ref{sec:asym_comp} by including 4-body decays for $\phi_H$ mediated by $\phi^3 H^*$.
In this section we illustrate models in which it's possible to replicate more closely the mechanism with a secondary 3-body decay channel, but that predict unstable dark matter.
After arguing about bounds coming from the lifetime of the DM in Section \ref{sec:tcbdecay}, we will go through the other two classes of models introduced in \ref{sec:tcbbreaking}. We anticipate one of the two is viable only by taking the scalar $\phi$ mass close to the Planck scale.
\subsection{Critical dimensionality of the DCb violating operator}\label{sec:tcbdecay}
As argued in Section \ref{sec:higherdim}, integrating out the heavy scalar can lead to higher dimensional operators mediating fast DCb decay.
By taking the couplings $y\gtrsim \mathcal{O}(10^{-1})$, $\mtcb\sim 30$ TeV and $M_\phi\gtrsim 10^{15}$ GeV, from ID bounds we get that that $d$, defined in equation \ref{eq:nr_lifetime}, must satisfy $d>6$. As shown in Section \ref{sec:doubleyuk}, this  corresponds to an operator obtained by integrating out a single $\phi$ line. 
Notice that $d$ is set by the powers of $M_\phi$ appearing in the graph, or in other words by the number of $\phi$ internal lines.
\subsection{Class 3 models: double Yukawa}\label{sec:doubleyuk}
In this class of model $\udb$ is broken by two Yukawa couplings that do not allow for a consistent assignment of such quantum number among the fields.
After integrating out the heavy scalar $\phi$, we generate operators that will mediate the decay of the DCb. Indeed, by hypothesis, we assumed that in the two Yukawas the dark color contraction are such to violate $\udb$. At the UV lagrangian level, when integrating out the scalar, we generate an effective operator that contains the "unconfined DCb" $\Psi^{\ndc}$ ($\epsilon$ contracted). Let's focus on each of the three possibilities for this scenario:
\begin{itemize}
    \item Mixed-Mixed. This case is only possible for $\ndc=2$, since the effective operator is $\epsilon_{ij}\Psi^i \Psi^j \psi \psi$. This implies that the dark color group is actually $\mathrm{SO}(3)$ (but with DCquarks in the spinorial rather than the fundamental). This possibility does not appear in the original ACDM models. Anyway, the operator of dimension 6 mediates a too fast decay for the DCb, assuming $M_\phi\leq M_{\mathrm{P}}$.
    \item Dark-Mixed. Integrating out $\phi$ gives the operator $\epsilon_{ijk}\Psi^i \Psi^j \Psi^k \psi$. This is possible only for $\ndc=3$, as can be also shown by $N$-ality considerations on the two Yukawa terms. This operator can mediate fast decay, since $\psi$ can radiate a SM gauge boson and make the DCb decay into SM kinematically allowed. 
    \item Dark-Dark. In this case, the effective operator contains only DCquarks, and is $\epsilon_{ijkl}\Psi^i \Psi^j \Psi^k \Psi^l$. This works only for $\ndc=4$. This term does not mediate a decay of a DCb directly. However, if one of the DCquarks involved is charged under SM, it's possible to radiate a SM gauge boson and decay into SM particles. Notice that the total asymmetry of the $\epsilon$ tensor forces different flavors of DCquarks in order to not have a vanishing effective term. This can also be seen at the UV level, since the only $\phi$ dark color representation that allows non-vanishing Yukawas with different contraction structure is the 6-dimensional antisymmetric representation (which is self conjugate). Since all GC models that have at least 4 flavors of DCquarks have one charged under SM, decay into SM is possible.
    Another possibilty for the DCb decay is to radiate a Higgs pair from the internal $\phi$ line. 
    Since the effective operator is effectively 6d, it mediates too fast decays, as shown in Section \ref{sec:tcbdecay}.
\end{itemize}
In conclusion, models in which $\udb$ breaking happens through 2 Yukawas with inconsistent $\udb$ charges assignments are phenomenologically viable only if $\phi$ is around the Planck scale. 
The presence of the additional fields of the GUT completion might even worsen the problem of stability of the DCb, because the new particles might generate dangerous operators at scales lower than $M_\phi$.
Since this is a model dependent issue, we will not further discuss such models.
\subsection{Class 2 models: "failed baryogenesis"}
In this class of models $\udb$ breaking happens through a potential term for the scalar $\phi$ and a mixed Yukawa between the DS and SM (unlike class 1 models in which the Yukawa was only involving DS fields).
The presence of the mixed Yukawa forces $\phi$ to be in the (anti)fundamental of $\sun$.
The DCb decays into SM but is long lived for heavy enough scalars. The difference with respect to the case in Section \ref{sec:doubleyuk} is that the scalar appears with a potential term $\propto \phi^2,\phi^3, \phi^4$ that violates $\phi$ number, and therefore also $\udb$ in conjunction with the mixed Yukawa.
As a consequence, the potential term forces additional $M_\phi$ suppression in the EFT operator due to more internal propagators, bringing the $d$ appearing in equation \ref{eq:nr_lifetime} up to 7. In this case bounds on the lifetime of the DCb can be satisfied for $M_\phi \gtrsim 10^{15}$ GeV, as low as allowed by the weak washout condition for $\mathcal{O}(0.1)$ couplings.
In particular, in this scenario, the $d$ of the operator in the various cases is the following:
\begin{itemize}
\item $\phi^2$ ("Majorana" mass for $\phi$): this lead to effectively 6d operators $\epsilon_{ij}\Psi^i\Psi^j \psi \psi$, therefore the presence of a Majorana mass term with a mixed Yukawa is disfavored. Therefore we will not consider theories in which is possible to write a Majorana mass term and a mixed Yukawa. Also, it's only possible in a $\mathrm{SU}(2)_{\mathrm{DC}}$ theory (because $\phi$ must belong to the fundamental gauge representation), which are not found in \cite{Antipin:2015xia}. 
\item $\phi^2 H^n$: this potential term leads to effective operators of the form $\Psi \Psi \psi \psi H^n$, with a $M_{\phi}^{4-n}$ suppression. Therefore they are safe from bounds for $n>1$. Again, this is possible only for $\mathrm{SU}(2)_{\mathrm{DC}}$ theories, which do not satisfy the constraints of the original ACDM models.
\item $\phi^3H^n$: this term generates effective operators of the form $\epsilon_{ijk}\Psi^i \Psi^j \Psi^k \psi^3  H^n$. The suppression is $M_\phi^{5+n}$. It's possible only for $\sut$ theories. In both cases, this lead to unstable but long-lived DM. Notice that these potential terms were found in the models with stable DM in Sections \ref{sec:phi3} and \ref{sec:phi3h} ($n=0,1$ respectively).
\item $\phi^4$: this term generates the effective operator $\epsilon_{ijkl}\Psi^i \Psi^j \Psi^k \Psi^l \psi^4$. The suppression is $M_\phi^8$. It's possible only for $\mathrm{SU}(4)_{\mathrm{DC}}$ theories. The DCb therefore is unstable, but long-lived to escape decay bounds as in the models of Section \ref{sec:othermodels}. This potential term is the same found in Section \ref{sec:phi4}.
\end{itemize}
As a concrete example of this class of models, we fix $G_{\mathrm{DC}}=\sut$ and the complex dark scalar $\phi$ in its antifundamental, while the dark quarks field content and SM quantum numbers are the left handed $Q=(3,2)_{1/6}$ and $D=(3,1)_{-1/3}$ and their right-handed conjugates (in principle there are many different possibilities, here we only pick one). 
The possible portals are:
\begin{equation}\label{eq:ex1_portals}
\mathcal{L}\supseteq y_q\phi \bar{q}_L Q_R + y_d\phi \bar{d}_R D_L + \mathrm{h.c.}\;,
\end{equation}

while the cubic term for the scalars is:
\begin{equation}\label{eq:ex1_cubic}
M \lambda\epsilon^{ABC}\epsilon^{ijk}\phi_{Ai}\phi_{Bj}\phi_{Ck}\;.
\end{equation}
where $ABC$ are flavor indices and $ijk$ the $\sut$ gauge indices, and $M$ is the scale of the cubic interaction. This term would vanish in the presence of a single flavor: that's why we need $3$ at least to make a non-zero cubic term. Notice that this type of asymmetric extension is rather general since it can be applied to any $\sun$ model listed in \cite{Antipin:2015xia} simply by pairing the DCquarks to the corresponding SM fermions. \\
As a side effect of the mixed Yukawa, the SM and DS asymmetries are related by an unbroken symmetry rotating the DCquark and SM fermion:

\begin{equation}
\label{eq:asymrel}
n_{\mathrm{SM}}=-n_{\mathrm{DS}}\;.
\end{equation}

Notice that now the Yukawa couplings in equation \ref{eq:ex1_portals} become $3\times 3$ matrices: there is enough room to have a physical CP violating phase. \\
Equation \eqref{eq:asymrel} implies the following relation between the dark and visible sector abundances:

\begin{equation}
    \Omega_{\text{B}} = \frac{m_p}{\mtcb}\Omega_{\text{DM}}
\end{equation}

so that for $\mathcal{O}$(1) TeV DM mass, assuming the correct DM relic density $\Omega_{\mathrm{DM}}$ is reproduced, the asymmetry generated from this process in the visible sector is only a subdominant fraction of the present-day observed SM asymmetry. Hence, despite being a viable model on his own, it cannot be a successful baryogenesis model unless the DM mass itself is in the GeV ballpark. In fact, this last possibility can be achieved by taking the model with the SM singlet $N$ as the only DCquark. Indeed, due to the absence of any charged state, it is possible in principle to take $\ldc$ close to the GeV scale and achieve a scenario where the asymmetry is correctly generated for both the SM and DM.
The fermionic field content of such model forces $\phi$ to carry hypercharge in order to have a non trivial Yukawa with the SM. So the potential term has to be $\phi^3 H^*$, which, together with the request of the existence of the Yukawa, determines the SM quantum number of $\phi$, and forces the Yukawa to be 
\begin{equation}
    \phi^\dagger q_L N \;,
\end{equation}
where $\phi$ has to carry also $\mathrm{SU}(3)_c$ and $\mathrm{SU}(2)_L$ charges.
Investigating this scenario is outside the scope of this work.\\
Apart from this special case, it is important to realize that portals like \eqref{eq:ex1_portals} do not necessarily spoil any asymmetry-generating mechanism in the SM, such as the usual thermal leptogenesis scenario \cite{Fukugita:1986hr}. Indeed, if the temperature at which the usual $B-L$ asymmetry is generated in the SM is much lower than the mass of the scalar, the asymmetry generated in the SM will not be transferred to the dark sector since the SM and the DS are not chemically coupled at such temperatures. Also, any initial $B-L$ asymmetry in the SM will be washed out by the SM-asymmetry generating mechanism, such as $\Delta L=2$ processes involving virtual and real Majorana neutrinos in the leptogenesis case. In other words, the asymmetry that is simultaneously generated from $\phi$ decay into SM is a byproduct of the mixed Yukawa which bears no consequences on the baryon asymmetry of the Universe. \\
A final comment for this class of models is the effect of the specific GUT realization on the DCb decay. The presence of the mixed Yukawa with the SM field forces $\phi$ to be either in the fundamental or antifundamental of $\sun$. In both cases, only the SM quantum numbers can prevent the existence of extra Yukawas that could make the model  effectively class 3, and therefore unviable. The same is true also when discussing the presence of higher dimensional operators that could mediate fast DCb decays.
Unlike the class 1 models presented in Section \ref{sec:buildingmodels}, $\sun$ invariance does not automatically protect the DCb from fast decay in presence of GUT partners. Given that the question depends on the particular ways in which the GUT completion is realized, we will not further analyze it.

\section{Phenomenology of Asymmetric ACDM models}
\label{sec:pheno}

In this Section we summarize the relevant phenomenology of Asymmetric ACDM. As we shall see, the different searches can discriminate between a symmetric and an asymmetric scenario (gravitational waves) or between an elementary or composite one (sizeable dipole moments from direct detection experiments), as well as signatures which are typical of composite asymmetric models (indirect detection from bound state formation). Finally, we also discuss the case of oscillations, predicted in one of the classes of models that we introduced, which can lead to present day residual annihilations. All in all, the combination of the various searches can single out the Asymmetric ACDM scenario, so that finding consistent UV completions is less a theoretical whim than the need to provide a model building base to a precise phenomenological picture.

\subsection{Gravitational waves}

\begin{figure}[htp!]
    \centering
    \includegraphics[width=0.65\textwidth]{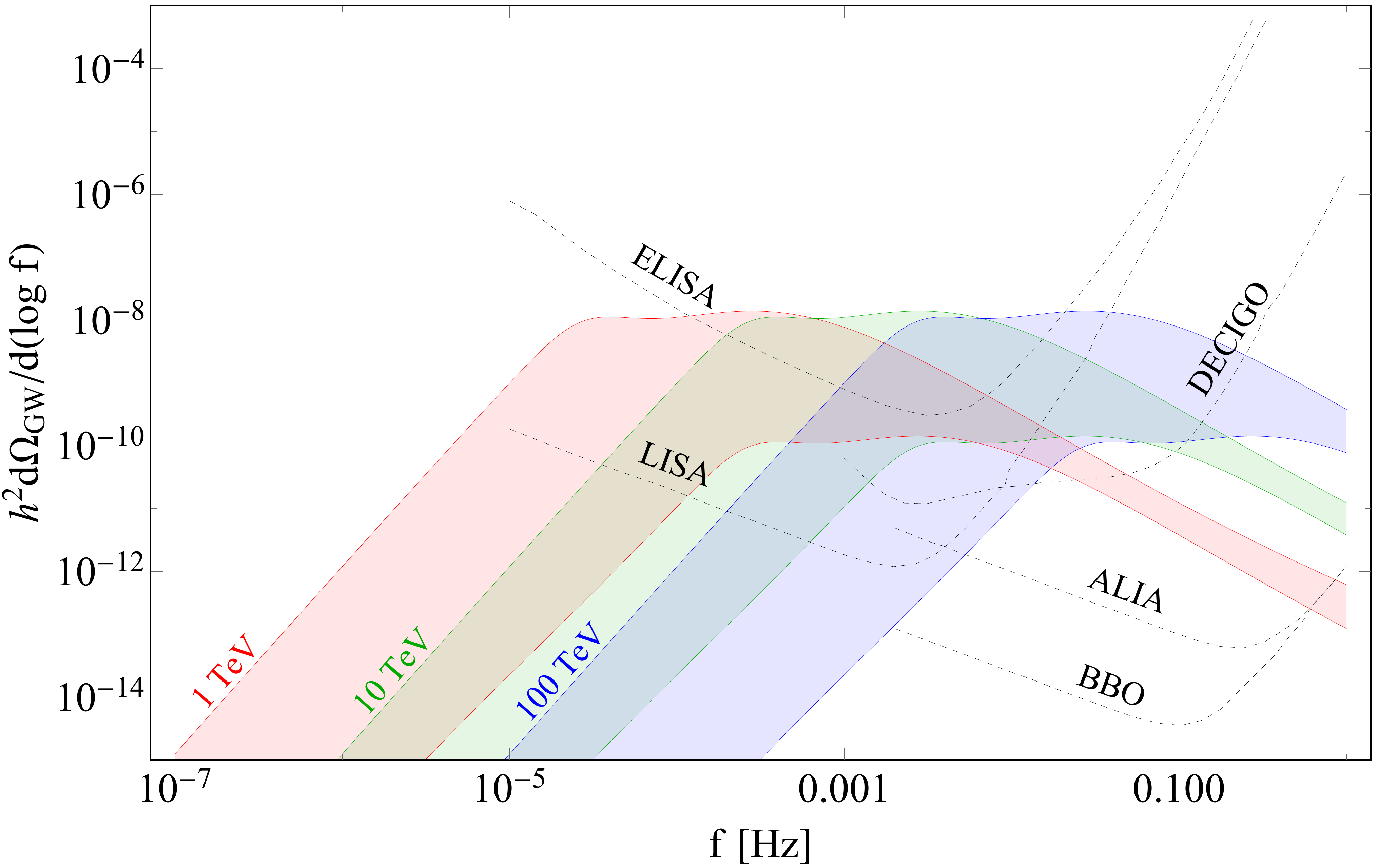}
    \caption{Spectra of GWs from a dark first order phase transition compared to projected reach of several future experiments. The upper and lower edges of the bands correspond to different nucleation rates $\beta=\mathcal{H}$ and $\beta= 10\mathcal{H}$, respectively, where $\mathcal{H}$ is the conformal Hubble rate at $T=T_*$.}
    \label{fig:gw}
\end{figure}

The observation of Gravitational Waves (GWs) originating from a dark first order phase transition is an exciting possibility since it can lead to observable signatures even from completely secluded dark sectors. In $\sun$ ACDM models with $n_f$ flavors, the confinement phase transition at $T_*\approx \ldc$ is first order if $3\leq n_f < 4\ndc$ (or $n_f=0$), where the second inequality is the requirement of asymptotic freedom. During the phase of nucleation, magnetohydrodynamics turbolences in the plasma induced by the expansion of the bubbles or collisions among the bubbles themselves are both sources of GWs. Following  \cite{Schwaller:2015tja}, in Figure \ref{fig:gw} we show the GW spectra for different confinement scales compared to the projected reach of future GW detectors \cite{Moore:2014lga}.

As we can see, because of the different position of the peaks, future interferometers could discriminate a symmetric from an asymmetric ACDM scenario. In particular, the former would be disfavored if the detected GW spectrum is compatible with a phase transition occurring at $T_*\lesssim 10$ TeV. 

\subsection{Direct Detection}

One of the features of ACDM models is that, despite its overall neutrality, the DM candidate has in general electrically charged components. In the case of fermionic DCb, for example, this leads to magnetic or electric dipole moment interactions of the DCb $\mathcal{B}$ with the photon generated by the following 5D operators:

\begin{equation}
    \Delta \mathcal{L} = \overline{\mathcal{B}}\sigma_{\mu\nu}(\mu_M+id_M\gamma^5)\mathcal{B} F^{\mu\nu}
\end{equation}

where the dipole moments are parametrized in terms of the gyromagnetic and gyroelectric factors as follows:

\begin{equation}
    \mu_M=g_M\frac{e}{2\mtcb}, \quad d_E=g_E\frac{e}{2\mtcb}.
\end{equation}

\begin{figure}[htp!]
     \centering
     \begin{subfigure}[b]{0.48\textwidth}
         \centering
         \includegraphics[width=\textwidth]{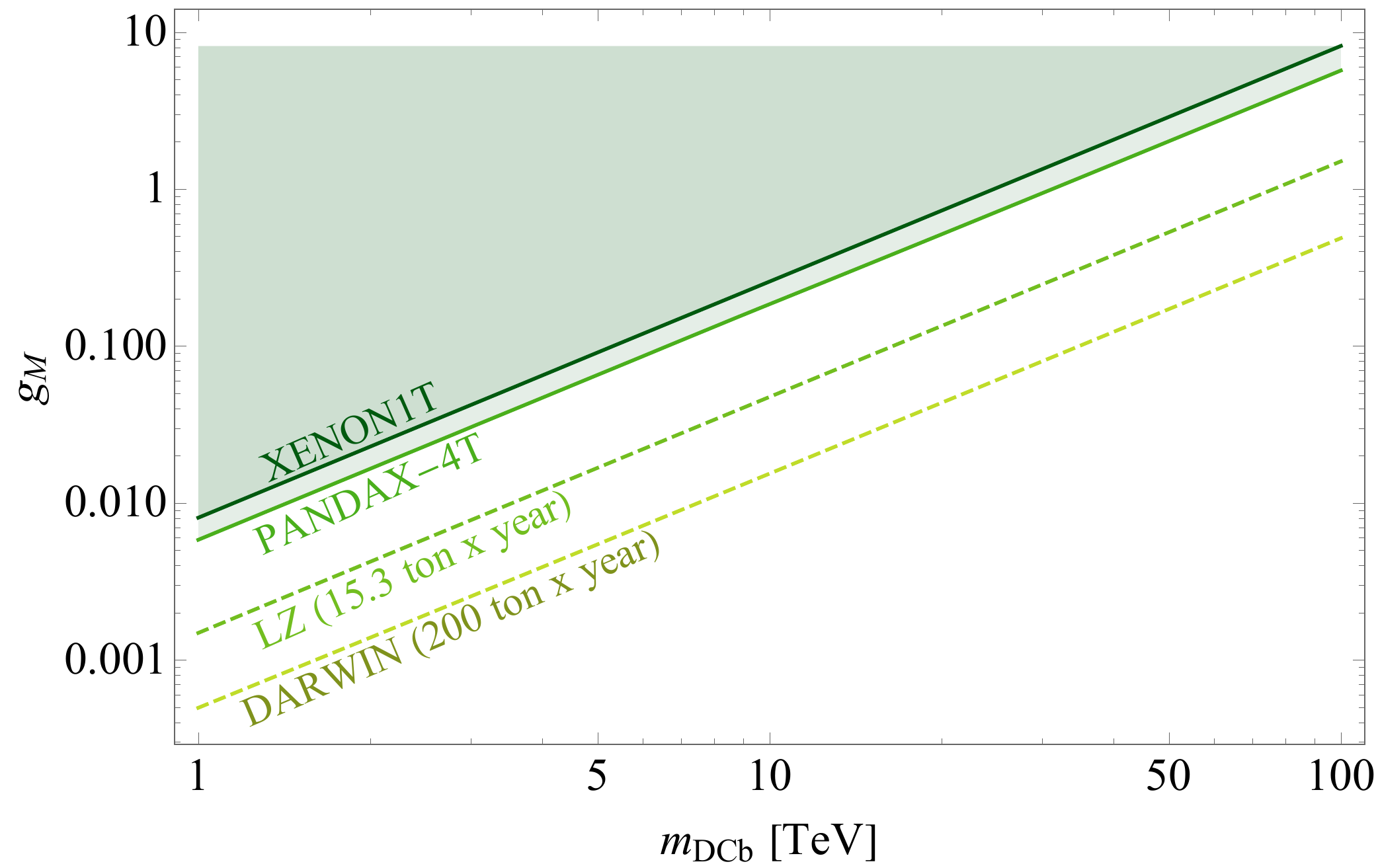}
         \caption{}
         \label{fig:mag}
     \end{subfigure}
     \hfill
     \begin{subfigure}[b]{0.48\textwidth}
         \centering
         \includegraphics[width=\textwidth]{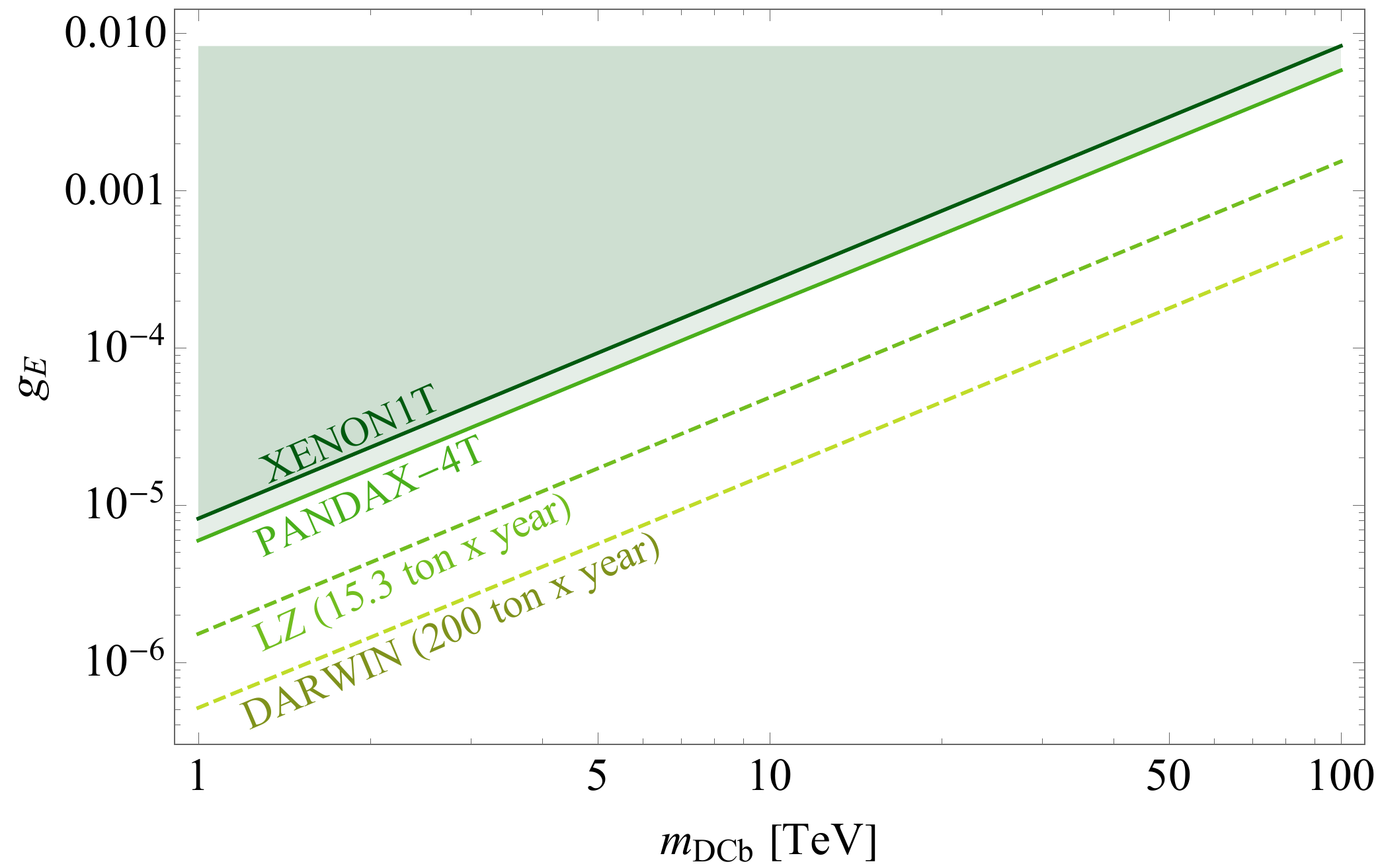}
         \caption{}
         \label{fig:ele}
     \end{subfigure}
        \caption{In dark green we show the exclusion regions on the gyromagnetic (left) and gyroelectric (right) factors from the XENON-1T \cite{XENON:2018voc} and PandaX-4T \cite{PandaX-4T:2021bab}. The dashed lines represent the projected exclusion limits from future LZ \cite{Mount:2017qzi} and DARWIN \cite{DARWIN:2016hyl}.}
        \label{fig:dipoles}
\end{figure}

In particular, from the strong dynamics of the dark sector a $\mathcal{O}(1)$ $g_M$ is expected, while the gyroelectric factor can be estimated as follows:

\begin{equation}
    g_E \sim \theta_{\mathrm{DC}} \frac{\min{m_\Psi}}{\mtcb}
\end{equation}

where a non-zero $\theta$-angle is necessary due to the otherwise CP-preserving dynamics of the dark sector. Dipole moments lead to direct detection cross-sections with peculiar dependencies on the nuclear recoil energy $E_R$\cite{Kavanagh:2018xeh}:

\begin{equation}
\begin{split}
    \frac{\mathrm{d}\sigma_M}{\mathrm{d}E_R}=&\frac{\mu_M^2\alpha_{\mathrm{em}}}{E_R}F_M^{pp}+\mathcal{O}\left(\frac{1}{v^2}\right)\\
    \frac{\mathrm{d}\sigma_E}{\mathrm{d}E_R}=&\frac{d_E^2\alpha_{\mathrm{em}}}{E_R v^2}F_M^{pp}
\end{split}
\end{equation}

where $F_M^{pp}$ is a nuclear form factor \cite{Fitzpatrick:2012ix}. In particular, the SI part of the magnetic dipole cross-section has a $1/E_R$ enhancement at low $E_R$, while the electric dipole cross section has an even stronger enhancement of $1/E_R v^2$. Following the matching and rescaling procedure outlined in \cite{Cirelli:2013ufw}, in Figure \ref{fig:dipoles} we show the constraints on the gyromagnetic and gyroelectric coupling from present and future direct detection experiments.

While $g_M\geq 1$ is already excluded by XENON1T and PANDAX-4T, future large exposure experiments like DARWIN can extend such exclusion to the entire range of masses typical of both symmetric and asymmetric ACDM. Similar conclusions hold for the the electric dipole moment, where $\theta_{\mathrm{DC}}=\mathcal{O}(1)$ is disfavored because it would imply very small masses for the constituent techniquarks, possibly at odds with collider bounds on \tcp. On a final note, the strong dynamics could make even higher dimensional operators like the charge radius operator $\overline{\mathcal{B}}\gamma_\mu \mathcal{B}\partial_\nu F^{\mu\nu}$, or $\overline{\mathcal{B}}i\partial_\mu \mathcal{B}\partial_\nu F^{\mu\nu}$ for scalar DM, more important than dipole interactions\cite{Kavanagh:2018xeh}, though it does not exhibit the $1/E_R$ enhancement. In the scalar case, in particular, the charge radius operator is the lowest dimensional operator that couples DM to the nuclei. Additional interactions make the previous bounds on dipole moments stronger.

\subsection{Indirect Detection from Bound State Formation (BSF)}

One weak side of typical ADM models is the absence of relevant indirect detection signatures due to lack of antiparticles. However, as pointed out by \cite{Mahbubani:2019pij}, this may not be true for models of \textit{composite} asymmetric DM thanks to the formation of dark nuclei. In such process, two DM particles bind into a deuteron-like state, emitting radiation with energy equal to the nuclear binding energy $E_B$, due to the small initial relative velocity. In the simplest case considered in \cite{Mahbubani:2019pij}, the DM belongs to an integer representation of SU(2)$_L$, like in Minimal Dark Matter (MDM) models \cite{Cirelli:2005uq}, and can be realized in the $\Psi=V$ model, where the dark baryon $V^3$ is a weak isotriplet. In this case, BSF proceeds through the emission of a photon via a magnetic dipole transition\footnote{Electric dipole transitions are also present but more suppressed \cite{Mahbubani:2020knq}.} mediated by the operator:

\begin{equation}
    \mathcal{L}_{\mathrm{mag}}=g_M \frac{e}{\mtcb}\overline{V^3}(\vec{\sigma}\cdot\vec{B})J_3 V^3
\end{equation}

where $\vec{B}$ is the magnetic field and $J_3$ the third isospin generator. Upon solving the Schroedinger equation:

\begin{equation}
    -\frac{\mathrm{d}^2u}{\mathrm{d}r^2}+V(r)u=m_{\mathrm{DCb}} v^2 u
\end{equation}

where $u$ is the reduced wave function describing the DM multiplet and $V(r)=V_{\mathrm{EW}}(r)+V_N(r)$ the potential containing both the electroweak and the nuclear contributions, the BSF cross-section is given by:

\begin{equation}
    \sigma_B v_{\mathrm{rel}}= 8 g_M^2 \frac{E_B^3}{m_{\mathrm{DCb}}^2}\left|\int\mathrm{d}r u_i^\dagger u_f\right|^2
\end{equation}

where $u_f$ is the reduced wave function describing the nucleus while $u_i$ the solution to the Schroedinger equation. The BSF cross-section is then compared to the bound by noticing that, while the DM number density in the DM halo is set by the mass, the energy of the emitted photon is given by $E_B$, so that:

\begin{equation}
    (\sigma_B v_{\mathrm{rel}})<2\left(\frac{m_{\mathrm{DCb}}}{E_B}\right)^2\left.(\sigma_{\mathrm{ann}} v_{\mathrm{rel}})\right|_{\mtcb=E_B}
\end{equation}

\begin{figure}[htp!]
    \centering
    \includegraphics[width=0.65\textwidth]{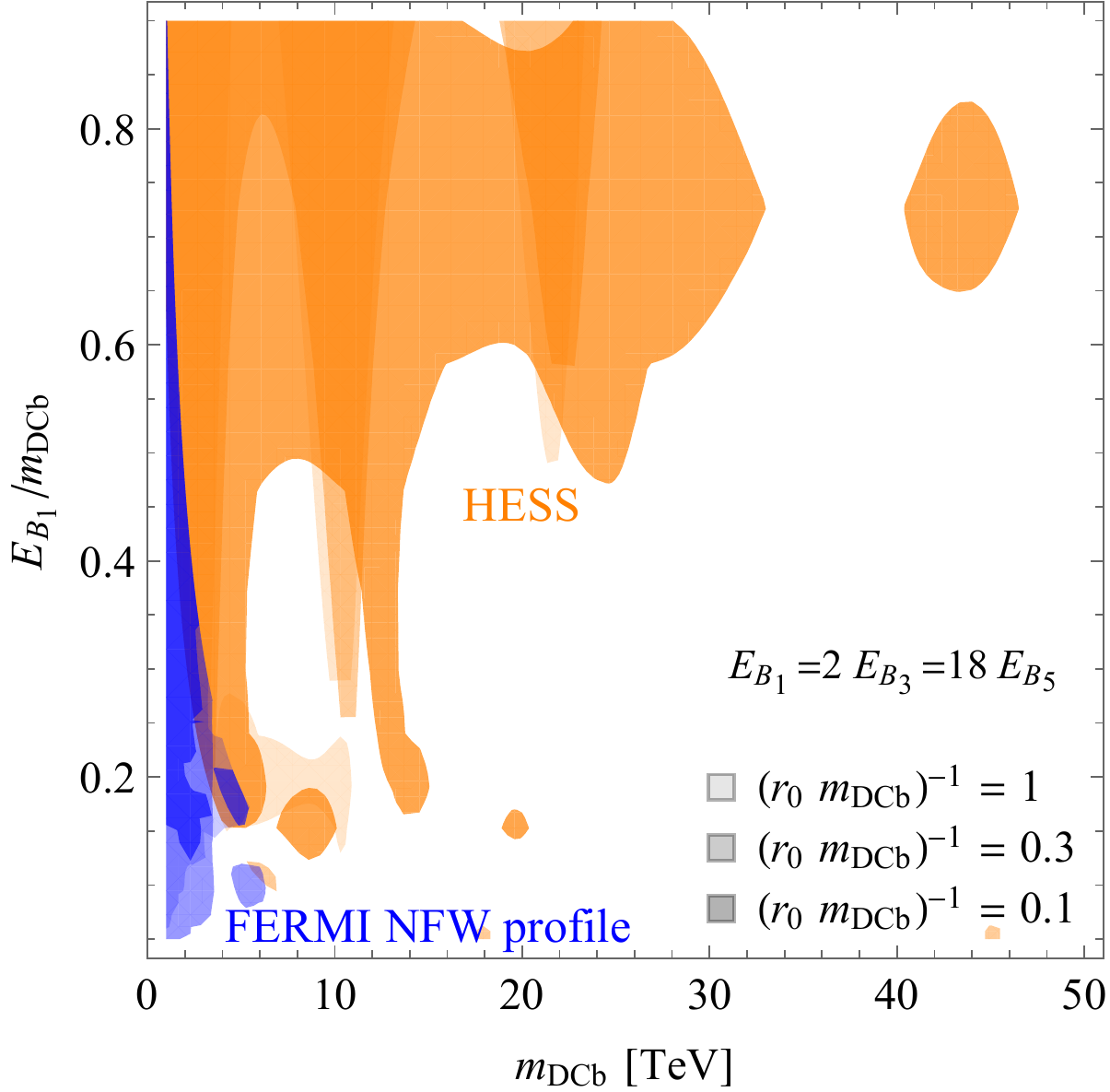}
    \caption{Binding energies of the deepest nuclear state, the isosinglet $B_1$ excluded by FERMI (blue) and HESS (orange) telescopes. Darker hues refer to larger nuclear radii in units of the DM mass. The $\gamma$-ray spectrum is obtained under the assumption that the nuclear potential can form a single shallow bound state for each isospin channel and by fixing the ratio of binding energies to $E_{B_1}=2E_{B_3}=18E_{B_5}$.}
    \label{fig:ID_bsf}
\end{figure}

Following \cite{Mahbubani:2019pij}, we assume that DM can only form a single shallow nucleus in each isospin channel, which in this case translates into the following isospin-spin configurations $1_0\oplus 3_1\oplus 5_0$. In Figure \ref{fig:ID_bsf} we then show the values of the binding energy of the deepest state (the isosinglet in our case) in order to produce a detectable signal for indirect detection from the FERMI \cite{Fermi-LAT:2015kyq} and HESS \cite{HESS:2018cbt} observation of the galactic center ($\gamma$-rays from dwarf galaxies lead to no appreciable bound). 

As we can see, BSF proves effective in producing a detectable signal in the low mass range $\mtcb\lesssim 35$ TeV, which is completely covered by HESS for deep enough bound states.\\ 
Another remarkable feature of this kind of dynamics is that, because of the selection rules on the magnetic dipole interactions, the nucleus that is initially formed is the isotriplet which later decays into the isosinglet state. Therefore, together with the photon emitted in the capture process, another monochromatic photon can be observed from the subsequent decay. This would represent a staggering signature pointing towards BSF in the dark sector.\\

Finally, apart from GC models containing $V$ in the DCq spectrum, the DM candidate is a complete SM singlet DCb, so that the typical EW potentials $V_{\mathrm{EW}}(r)$ as computed, for example, in MDM models are now absent. However, the EW interactions may be replaced by Yukawa interactions among the DCq, which are in general present in order to break unwanted species symmetries. These Yukawa's may generate analogous interactions between the DM and another DCb, leading to a similar dynamics with respect to that considered in this Section. Such possibility will be investigated in a future work.

\subsection{Oscillations}\label{sec:pheno_osc}
A feature of the Class 1 models is that the DM itself can oscillate, thanks to the Majorana mass of Eq. \ref{eq:IR_mass} induced by the heavy scalars. Since the scalars are heavy, both for naturalness and to avoid washing out the asymmetry, they are hardly detectable at current or future collider experiments. Testing the oscillation can therefore be a probe of the mass of the scalar.
The effect of the oscillation is to regenerate a symmetric component of the DM.
As a consequence, DCb-$\overline{\mathrm{DCb}}$ residual annihilation is enhanced, and possibly can even recouple.
As anticipated in Sec. \ref{sec:tcboscillation}, this suggests the possibility to probe the oscillation rate, and therefore $M_\phi$, via ID experiments \cite{PhysRevLett.108.011301}.
In general, we expect the DCb-$\overline{\mathrm{DCb}}$ annihilations to produce a number of DC$\pi$, similarly to what happen in the SM with proton-antiproton annihilations \cite{Orfanidis:1973ix}. Such channels were analyzed in \cite{Ibe_2020}, although in different models of composite DM.
For example, in the $N_{\mathrm{DC}}=3$, $\Psi=V$ model we expect the DCb to be a $\mathrm{SU}(2)_L$ triplet, and therefore the DCb-$\overline{\mathrm{DCb}}$ system can be decomposed in its different isospin components. Each of these components can annihilate in a different number of DC$\pi$, consistently with $\mathrm{SU}(2)_L$ invariance and G-parity (the conservation stems from the fact that the annihilation process proceeds through the $\sun$ interaction, which conserves the other quantum numbers).
Each of the final state DC$\pi$ will then decay in the SM, either via the chiral anomaly in photons, or via Higgs and other SM gauge bosons (if there are extra DCquarks and Higgs portals).
A precise spectrum computation of such lines is outside the scope of the current work.
To give a crude estimate of the feasibility to probe the residual annihilations, we assume that the number of DC$\pi$ in all relevant final states is $\mathcal{O}(1)$. Given that such states can then give rise to lines, we expect that the $\gamma$-ray spectrum will be peaked roughly at $\mtcb/2$, although it will present a spread due to the multi-body nature of the DCb annihilation processes.
Current experiments like HESS are sensitive to $\gamma$-rays in the multi-TeV range, and are able to exclude annihilation cross sections of order $\langle \sigma v \rangle\sim 10^{-22}\div 10^{-24} \; \mathrm{cm}^3/ \mathrm{s}$ \cite{HESS:2020zwn}.
Although such exclusions are taken for monochromatic annihilation spectra, we will study two benchmark values of $\langle \sigma v \rangle$, to get a crude picture of whether or not it's possible at current experiments to probe the oscillations.

\begin{figure}
    \centering
    \includegraphics[width=0.65\textwidth]{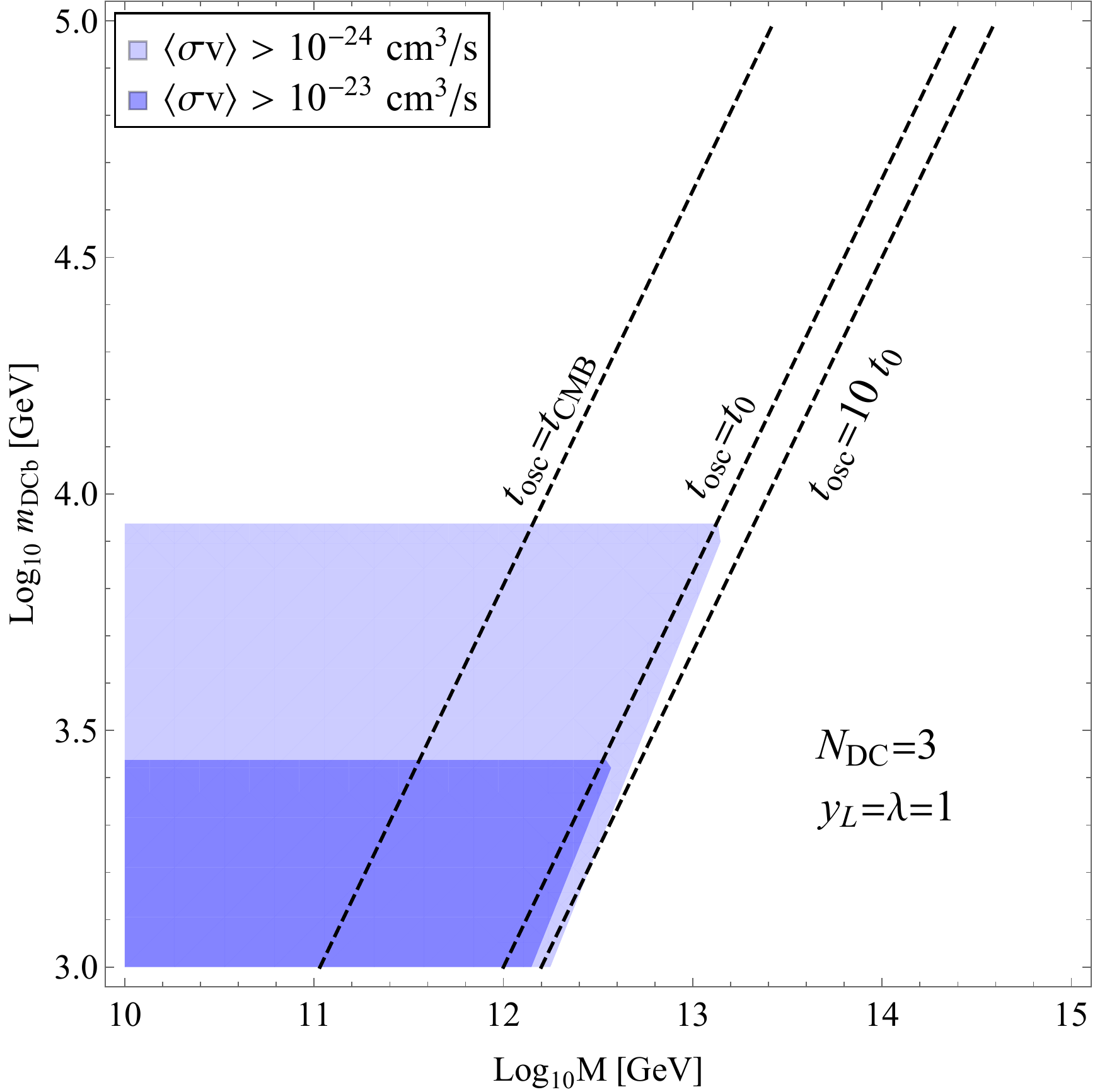}
    \caption{$\gamma$-ray bounds from dwarf galaxies in the $M_\phi$-$\mtcb$ plane for two different values of the cross section. Bounds are computed by assuming $\lambda=y_L=1$. Shadowed regions can be potentially tested and excluded.}
    \label{fig:osc_bound}
\end{figure}

To recast the bound, we notice that in the oscillating case, for large oscillation periods $t_\mathrm{osc}$, the ratio between $\overline{\mathrm{DCb}}$ and DCbs is roughly $(t/t_\mathrm{osc})^2$, given that the probability of conversion is proportional to $\sin^2(t/t_\mathrm{osc})$
Therefore the quantity that enter the bound is:
\begin{equation}\label{eq:residual_anni}
  \left(\frac{t}{t_{\mathrm{osc}}}\right)^2\langle \sigma v\rangle \approx \left(\frac{\Delta m}{\hub}\right)^2 \frac{25}{\mtcb ^2} \;,
\end{equation}
where $\Delta m$ is the DCb mass splitting of Eq. \ref{eq:mass_splitting}.
This kind of estimate only works if the residual annihilation and elastic scattering process do not make the oscillation process lose its coherence, as suggested in \cite{Tulin:2012re, Cirelli:2011ac}. In our models this is possible because the elastic scattering with thermal bath particles can be suppressed via mass splitting of the DCb components (due for example to EW mass splittings in the $V$ model) or be mediated by higher dimensional operators involving the Higgs (if $N^3$ is the DCb for example), while annihilations and scatterings with other DCbs are always Boltzmann suppressed in a fully asymmetric regime. So we will neglect such processes and use Eq. \ref{eq:residual_anni}. 
We plotted the results in figure \ref{fig:osc_bound}.
The horizontal lines are due to the fact that Eq. \ref{eq:residual_anni} only works for $t\ll t_{\mathrm{osc}}$, since for $t$ large enough the densities of particles and antiparticles will reach and oscillate around half the initial particle density. In this case the bound on the cross section (and therefore on $\mtcb$) saturates since no larger value for $n_{\mathrm{DCb}}n_{\overline{\mathrm{DCb}}}$ can be reached.
We checked that for $M_\phi\gtrsim 10^{10}$ GeV oscillations start only at temperatures too low for the residual annihilations to recouple and change the total DM abundance, therefore there is no thermal recoupling of the annihilations.
For $M_\phi$ in the range of the plot in Fig. \ref{fig:osc_bound}, the weak washout condition can be fulfilled by taking the Yukawa of the heavy scalar $y_H$ to be around $10^{-3}$.
Notice that only a thin portion of the plane, around $M_\phi \lesssim 10^{13}$ GeV has a cross section in range of current $\gamma$-ray experiments. However, for lighter $\phi$, oscillations can start even at CMB or before, as shown by the dashed lines. In this case, the asymmetry can be totally washed-out if it's not stored in dark nuclei. CMB bounds could then in principle be studied for such models, possibly leading to stronger bounds.
To conclude, in our models the region $10^{12}\;\mathrm{GeV}\lesssim M_\phi\lesssim 10^{13}\;\mathrm{GeV}$ can be probed by studying residual annihilations, while for $M_\phi \lesssim 10^{12}$ GeV bounds from CMB could be applied. For heavier $M_\phi$ no bound from oscillations arises.
Another interesting fact to test the oscillations is that the asymmetry in the DCbs bound in dark nuclei is preserved, as argued in Section \ref{sec:tcboscillation}. This could lead to signatures of a dark nucleus "annihilating" with a $\overline{\mathrm{DCb}}$, and possibly to spectral shapes which are unique to our models.
\section{Conclusions}\label{sec:conclusion}
We have built a class of possible minimal UV completions to the Accidental Composite Dark Matter models that can produce the correct amount of asymmetry. In order to do so, we simply added two flavors of a heavy scalar $\phi$, with mass around the cutoff of the full theory $\lcut$. The scale $\lcut$ can be taken to be the Planck scale $M_{\mathrm{P}}$, the GUT scale (which is a natural cutoff of ACDM models), or as low as $M_\phi\simeq 10^{10}$ GeV, provided the additional fields at the cutoff do not introduce new non-trivial $\sun$ representations. Below the mass of the heavy scalars, the model behaves like the original ACDM models with a non-zero initial asymmetry. The DM candidate is a dark baryon, and with our choice of gauge representations it is accidentally stable in the IR theory. The mechanism we provided can be easily adapted to all the golden class ACDM models, even allowing the possibility for generic $\mathrm{SU}(5)$ GUT completions and asymmetry generation in the visible sector via, for example, thermal leptogenesis. The symmetric component of DM is eliminated thanks to non-perturbative annihilations below $\ldc$ due to residual dark color interactions, provided $\mtcb \lesssim 75 \,\mathrm{TeV}$, without the need of having new dark forces. This lower even further the scale of the DM mass (and of the typical resonances of the dark sector) with respect to the original symmetric ACDM models. The asymmetry generation mechanism can produce enough asymmetry for $\mtcb$ in this range. The choice of the coupling of the UV sector responsible for the asymmetry generation can be made natural (or at most fine-tuned at the percent level).
We have also built models in which the DM candidate is unstable but long-lived enough to satisfy current experimental bounds.\\
Future directions of this work include the exploration of other mechanisms that can give a common explanation to the asymmetry of the DS and the visible sector. The main obstruction to this kind of construction is that collider constraints force $\ldc$ to be larger than at least the TeV scale. This implies a natural hierarchy between the asymmetries in the two sectors, that can hardly be achieved through usual cogenesis mechanism realized via renormalizable portals\footnote{Unless one consider hierarchical couplings in the two sectors. This simply moves the problem from the hierarchy of the asymmetries to the hierarchy of the couplings.}. Possible solutions to this problem could be found exploiting peculiarities of the strong dynamics \cite{Asadi:2021yml}, and the dynamical generation of the $\ldc$ scale.
\section*{Acknowledgments}
The authors want to thank Michele Redi and Roberto Contino for early participation in the work and for useful comments and insights, Arnab Dasgupta for interesting discussions, and Paolo Panci for useful clarifications on direct detecion. OP is supported by the Samsung Science and Technology Foundation under Grant No. SSTF-BA1602-04 and National Research Foundation of Korea under Grant Number 2018R1A2B6007000. MC is partially supported by the PRIN 2017L5W2PT. All Feynman diagrams were created using TikZ-Feynman LateX package~\cite{Ellis:2016jkw}.

\appendix

\section{Complete asymmetry parameter}\label{app_asym}

In order to compute the asymmetry \eqref{asym_bm}, we first need to evaluate:

\begin{equation}
    \delta\equiv |\mathcal{M}|^2-|\overline{\mathcal{M}}|^2 =-4(\mathcal{I}_H + \mathcal{I}_L)\;,
\end{equation}

where $\mathcal{I}_H$ ($\mathcal{I}_L$) collects the contributions coming from the interference of the loop diagrams in Figure \ref{fig:threebody} with the tree-level one with $\phi_H$ ($\phi_L$) in the internal line. They are given, respectively, by:


\begin{equation}\label{eq:interference_H}
\begin{split}
    \mathcal{I}_H=&\frac{\Im[\lambda_{LHH}^*\lambda_{HLL}y_L^*y_H]}{x-1}\left(\frac{\alpha^2}{\alpha^2-1}\frac{|y_H|^2}{16\pi}\frac{1}{x-1}+4|y_H|^2\mathcal{D}_1^{HL}+2|y_L|^2\mathcal{D}_2^{LL}\right)\\
    &+\frac{\Im[\lambda_{LHH}^*\lambda_{LLL}y_L^{*2}y_H^2]}{x-1}\left(\frac{\alpha^2}{\alpha^2-1}\frac{\alpha^2}{16\pi(\alpha^2 x-1)}+2\mathcal{D}_1^{LL}\right)\\
    &+\Im[\lambda_{HLL}^*\lambda_{LLL}y_L^*y_H]\frac{\alpha^2}{\alpha^2-1}\frac{|y_L|^2\alpha^4}{16\pi(\alpha^2 x-1)^2}-\Im[\lambda_{LHH}\lambda_{HHH}^*y_L^*y_H]\frac{2|y_H|^2\mathcal{D}_2^{HH}}{x-1}
    \end{split}
\end{equation}


\begin{equation}\label{eq:interference_L}
\begin{split}
    \mathcal{I}_L=&\frac{\Im[\lambda_{LHH}^*\lambda_{HLL}y_L^*y_H]}{\alpha^2x-1}\left(\frac{1}{\alpha^2-1}\frac{|y_L|^2}{16\pi}\frac{\alpha^2}{\alpha^2x-1}-2|y_H|^2\mathcal{D}_1^{HH}-4|y_L|^2\mathcal{D}_2^{HL}\right)\\
    &+\frac{\Im[\lambda_{HLL}\lambda_{HHH}^*y_L^{*2}y_H^{2}]\alpha^2}{16\pi(\alpha^2 x-1)(x-1)}\frac{1}{\alpha^2-1}+\frac{2\Im[\lambda_{HLL}^*\lambda_{LLL}y_L^{*}y_H]|y_L|^2\alpha^2}{\alpha^2x-1}\mathcal{D}_2^{LL}\\
    &+\Im[\lambda_{LHH}\lambda_{HHH}^*y_Hy_L^*]\frac{1}{\alpha^2-1}\frac{|y_H|^2}{16\pi( x-1)^2}
    \end{split}
\end{equation}



where $x\equiv\frac{(p^\mu_{N1}+p^\mu_{N2})^2}{M_H^2}$, $p_{N1}$ and $p_{N2}$ being the momenta of the final fermions, and $\alpha=\frac{M_H}{M_L}$. The functions $\mathcal{D}_{1,2}^{ij}\equiv\mathcal{D}_{1,2}^{ij}\left(\frac{p_{N1}}{M_H},\frac{p_{N1}}{M_H},\frac{M_L}{M_H}\right)$ are the imaginary parts of the diagrams \ref{fig:loop_4} and \ref{fig:loop_5}, respectively, and are given by:

\begin{equation}
    \mathcal{D}_1^{ij}=\Im\left[\int\frac{\mathrm{d}^4l}{(2\pi)^4}\frac{\Tr[\slashed{p}_{N1}\slashed{p}_{N2}\slashed{l}(\slashed{p}_H - \slashed{l} )]}{l^2(p_H-l)^2((p_H-l-p_{N2})^2-M_i^2)((l-p_{N1})^2-M_j^2)}\right]
\end{equation}
\begin{equation}
    \mathcal{D}_2^{ij}=\Im\left[\int\frac{\mathrm{d}^4l}{(2\pi)^4}\frac{\Tr[\slashed{p}_{N1}\slashed{p}_{N2}(\slashed{l}-\slashed{p}_{N1})(\slashed{p}_H - \slashed{l} - \slashed{p}_{N2})]}{(l^2-M_i^2)((p_H-l)^2-M_j^2)(p_H-l-p_{N2})^2(l-p_{N1})^2}\right]
\end{equation}

where $p_H$ is the 4-momentum of the initial $\phi_H$ scalar, and $i,j$ run over the scalar flavors. These integrals have been evaluated by means of \texttt{Package-X}, setting $m_N=0$. Despite the different combination of couplings in the above interference terms, there are only three independent phases, as expected from \eqref{eq:int_lag}. Indeed, if we define:

\begin{equation}
    \text{arg}[\lambda_{LHH}^*\lambda_{HLL}y_L^*y_H]\equiv \theta_1, \quad \text{arg}[\lambda_{LHH}^*\lambda_{LLL}y_L^{*2}y_H^2] \equiv \theta_2, \quad \text{arg}[\lambda_{LHH}\lambda_{HHH}^*y_L^*y_H]\equiv \theta_3
\end{equation}

then the phases in the remaining combinations of couplings are given by:

\begin{equation}
    \text{arg}[\lambda_{HLL}^*\lambda_{LLL}y_L^*y_H]=\theta_2-\theta_1, \quad \text{arg}[\lambda_{HLL}\lambda_{HHH}^*y_L^{*2}y_H^{2}] = \theta_1+\theta_3
\end{equation}

In equations \ref{eq:interference_H}, \ref{eq:interference_L} we have omitted the color factors to avoid cluttering. To properly account for them, each function $\mathcal{D}^{ij}_{1,2}$ (related to box diagrams) must be multiplied by $C_{\mathrm{DC}}'=-27$, while the other terms (related to bubble diagrams) by $C_{\mathrm{DC}}=24$.
Once we plug the previous expressions into \eqref{asym_bm} and perform the integral over phase space, we get the asymmetry generated in our benchmark model with the complete set of parameters. Instead, if we set $\lambda_{LLL}=\lambda_{HLL}=0$ we recover \eqref{eq:dasgupta_factor}.

\section{Golden Class models}\label{sec:goldenclass}
In GC models the fermions are taken to be in the fundamental representation of $\sun$, and to be vector-like representations under the SM. In particular, it's assumed that the SM representations must be in $\mathrm{SU}(5)$ fragments. In table \ref{tab:tcquarks} we list the possible DCquarks out of which GC models are built.
\begin{table}[htp!]
    \centering
    \begin{tabular}{|c|c|c|c|c|c|}
    \hline
    $\mathrm{SU}(5)$  & $\mathrm{SU}(3)_c$ & $\mathrm{SU}(2)_L$ & $\mathrm{U}(1)_Y$ & charge & name\\
    \hline
    1 & 1 & 1 & 0 & 0 & $N$\\
    \hline
    $\bar{5}$ & $\bar{3}$ & 1 & 1/3 & 1/3 & $D$\\
     & 1 & 2 & -1/2 & 0,1 & $L$\\
    \hline
    10 & $\bar{3}$ & 1 & -2/3 & -2/3 & $U$\\
     & 1 & 1 & 1 & 1 & $E$\\
     & 3 & 2 & 1/6 & 2/3,-1/3 & $Q$\\
    \hline
    15 & 3 & 2 & 1/6 & 2/3, -1/3 & $Q$\\
     & 1 & 3 & 1 & 0,1,2 & $T$\\
     & 6 & 1 & -2/3 & -2/3 & $S$\\
    \hline
    24 & 1 & 3 & 0 & -1,0,1 & $V$\\
     & 8 & 1 & 0 & 0 & $G$\\
     & $\bar{3}$ & 2 & 5/6 & 4/3,1/3 & $X$\\
     & 1 & 1 & 0 & 0 & $N$\\
    \hline
    \end{tabular}
    \caption{List of possible DCquarks. Tilded DCquarks have same $\sun$ representation (\emph{i.e.} the fundamental), but conjugate SM representations with respect to the untilded counterparts.}
    \label{tab:tcquarks}
\end{table}

In Table \ref{tab:gc_models} we list the various golden class models identified in the original work \cite{Antipin:2015xia}.

\begin{table}[htp!]
    \centering
    \begin{tabular}{|c|c|c|}
        \hline
        DCq content & Allowed $\ndc$ & DM candidates\\
        \hline
        \multicolumn{3}{|l|}{$N_{\mathrm{DF}}=3$ } \\
        \hline
        $\Psi = V$ & 3 & $VVV$ =3\\
        $\Psi=N\oplus L$ & 3,...,14 & $N^{\ndc*}$\\
        \hline
        \multicolumn{3}{|l|}{$N_{\mathrm{DF}}=4$ } \\
        \hline
        $\Psi = V\oplus N$ & 3 & $VVV,\, VNN$ =3, $VVN$ =1\\
        $\Psi=N\oplus L\oplus \tilde{E}$ & 3,4,5 & $N^{\ndc*}$=1\\
        \hline
        \multicolumn{3}{|l|}{$N_{\mathrm{DF}}=5$ } \\
        \hline
        $\Psi = V\oplus L$ & 3 & $VVV$ =3\\
        $\Psi=N\oplus L\oplus \tilde{L}$ & 3 & $NL\tilde{L}$=1\\
        $=$ & 4 & $NNL\tilde{L},\, L\tilde{L}L\tilde{L}$=1\\
        \hline
        \multicolumn{3}{|l|}{$N_{\mathrm{DF}}=6$ } \\
        \hline
        $\Psi = V\oplus L\oplus N$ & 3 & $VVV,\, VNN$ =3, $VVN$=1\\
        $\Psi=V\oplus L\oplus \tilde{E}$ & 3 & $VVV$ =3\\
        $N\oplus L \oplus \tilde{L} \oplus \tilde{E}$ & 3 & $NL\tilde{L},\, \tilde{L}\tilde{L}\tilde{E}$=1\\
        $=$ & 4 & $NNL\tilde{L},\,  L\tilde{L}L\tilde{L},\, N\tilde{E}\tilde{L}\tilde{L}$=1\\
        \hline
        \multicolumn{3}{|l|}{$N_{\mathrm{DF}}=7$ } \\
        \hline
         $\Psi=L\oplus \tilde{L}\oplus E\oplus \tilde{E}\oplus N$ & 3 & $LLE,\, \tilde{L}\tilde{L}\tilde{E},\, L\tilde{L}N,\, E\tilde{E}N=1$\\
         $\Psi= N\oplus L\oplus \tilde{E}\oplus V$ & 3 & $VVV,\, VNN=3$, $VVN=1$\\
         \hline\hline
        \multicolumn{3}{|l|}{$N_{\mathrm{DF}}=9$ } \\
        \hline
        $\Psi=Q\oplus \tilde{D}$ & 3 & $QQ\tilde{D}=1$\\
        \hline
        \multicolumn{3}{|l|}{$N_{\mathrm{DF}}=12$ } \\
        \hline
        $\Psi=Q\oplus \tilde{D}\oplus\tilde{U}$ & 3 &  $QQ\tilde{D},\, \tilde{D}\tilde{D}\tilde{U}=1$\\
        \hline
    \end{tabular}
    \caption{$\mathrm{SU}(N)_{DC}$ golden-class models as classified in \cite{Antipin:2015xia}. For each model we specify the allowed number of dark colors which guarantee the perturbativity of the SM gauge group up to $M_{\mathrm{P}}$, and the DM DCb candidate with the corresponding $\mathrm{SU}(2)_L$ representation. A $*$ denotes a higher spin representation.}
    \label{tab:gc_models}
\end{table}

\section{Silver Class models}\label{sec:silverclass}
Silver Class models predict accidentally stable DC$\pi$, with unwanted charges under SM. This is due to the fact that the field content does not allow the presence of Yukawa with the SM Higgs that breaks the species number\footnote{We neglect the possibility of breaking  such symmetry using operators built with GUT partners since this is an issue dependent on the completion.}. In addition to these stable DC$\pi$s, there is a neutral DCb that is also accidentally stable thanks to the usual $\udb$. The introduction of the dark scalar $\phi$ could be used to both break $\udb$ and the unwanted species number, effectively "goldenizing" the model.
This is similar to what happens in the golden class $V\oplus N$ model, in which $\phi$ couplings can break residual species numbers.
We will show that $\sun$ SC models cannot be asymmetrized by implementing the benchmark model (or one of its variations): the dangerous \tcp must decay via scalar exchange before the formation of the earliest known structure, \emph{i.e.} before Big Bang Nucleosynthesis (BBN), $\tau_{\mathrm{BBN}}\simeq 1.5\times 10^{24} \; \mathrm{GeV}^{-1}$. This request is in contrast with the weak washout condition.
SC $\mathrm{SO}(N)_{\mathrm{DC}}$ models instead suffer the problem of containing only real candidates, and therefore cannot be asymmetrized.
\subsection{$\mathrm{SU}(N)_{\mathrm{DC}}$ models}

By energetic considerations, we expect the asymmetry to be shared by DC$\pi$s and DCbs, so that non-perturbative annihilations cannot annihilate entirely the population of DC$\pi$. Hence DC$\pi$ are forced to decay, but this is in contrast with indirect detection bounds. Indeed:
\begin{itemize}
    \item The ingredients to allow asymmetrization of a sufficiently long-lived DCb are, as in the GC model, the presence of a potential term for $\phi$ in conjunction with Yukawa interactions involving $\phi$ and the DCquarks.
    \item The breaking of the species symmetry stabilizing the \tcp is due to the presence of one of the Yukawa couplings between $\phi$ and the DCquarks.
    \item Since the decaying DC$\pi$ are SM charged, they must decay before BBN to avoid injecting extra energy in the SM fields. This argument relies on the fact the a non-negligible fraction of DM is stored in DC$\pi$. Indeed, we expect such fraction to be $\mathcal{O}\left(\sqrt{m_\Psi/\ldc}\right)$.
    \item This sets the bound $\Gamma_{\mathrm{DC}\pi}\gg \tau_{\mathrm{BBN}}^{-1}\simeq 6.6\times 10^{-25}$ GeV.
    \item By assumption, the minimum dimensionality of the effective operator mediating \tcp decay obtained by integrating out the heavy $\phi$ is 6. Indeed  in SC models no 5d operator mediating \tcp decay can be generated due to gauge invariance. 
    \item The upper bound for the \tcp decay width is: $\Gamma_{\mathrm{DC}\pi} \simeq \frac{y^4}{8\pi}\frac{M_{\mathrm{DC}\pi}^5}{M_\phi^4}$.
    \item By taking $\mathcal{O}(1)$ couplings and taking $M_{\mathrm{DC}\pi}\simeq 1$ TeV (to avoid collider bounds), we get $M_\phi \ll 10^{10}$ GeV.
    \item Such a mass for $\phi$ is outside the weak-washout regime, and the asymmetry generated by its out-of-equilibrium decay will be washed-out by inverse decay processes. Also, it suffers from a too large Boltzmann suppression, as remarked in Section \ref{sec:anotherestimate}.
\end{itemize}
Therefore our mechanism does not work to asymmetrize the SC models.
\subsection{$\mathrm{SO}(N)_{\mathrm{DC}}$ models}
The list presented in the Appendix of \cite{Antipin:2015xia} all the DCb DM candidates are self-conjugate: the gauge representation of the DCquark is self conjugate, and additionally, the DCb does not carry any species number needed to differentiate it from its antiparticles. Therefore it makes no sense to make these models asymmetric.

\bibliographystyle{JHEP}

\bibliography{bibliography}

\end{document}